\documentclass[12pt]{article}

\usepackage[top=1.2in, bottom=1.2in, left=0.75in, right=0.75in]{geometry}

\usepackage[sort&compress,authoryear]{natbib}
\usepackage{BOONDOX-cal}
\usepackage{epsfig, color}
\usepackage{graphicx}
\usepackage{amsmath,amsthm,amsfonts,amssymb,bm}
\usepackage[normalem]{ulem}
\usepackage{cancel}
\usepackage{authblk}
\usepackage[figuresright]{rotating}

\usepackage{makeidx}
\usepackage{listings}
\usepackage{algpseudocode}
\usepackage[colorlinks,
linkcolor=black,
anchorcolor=black,
citecolor=black
]{hyperref}
\usepackage{enumerate}

\usepackage{rotating}
\usepackage{graphicx}
\usepackage{tikz}
\usepackage{multirow}
\usepackage{booktabs}
\usepackage{threeparttable}
\usepackage{subfigure}
\usepackage{floatrow}
\usepackage{placeins}
\algrenewcommand\algorithmicrequire{\textbf{Input:}}
\algrenewcommand\algorithmicensure{\textbf{Output:}}
\newcounter{algorithm}
\newenvironment{algorithmblock}[1]{%
	\refstepcounter{algorithm}%
	\par\medskip
	\begin{center}
	\textbf{Algorithm \thealgorithm:} #1
	\end{center}
	\small
}{%
	\normalsize\par\medskip
}
\newtheorem{remark}{Remark}[section]

\newtheorem{assumption}{\hspace{6mm}Assumption}[section]
\newtheorem{theorem}{\hspace{6mm}Theorem}[section]
\newtheorem{proposition}{\hspace{6mm}Proposition}[section]

\newtheorem{definition}{\hspace{6mm}Definition}[section]
\newcommand{\1}{\mathbf{1}}
\makeatletter
\newcommand{\manualtaglabel}[2]{\phantomsection\def\@currentlabel{#2}\label{#1}}
\makeatother

\def\be{\begin{equation}}
\def\ee{\end{equation}}

\allowdisplaybreaks[4]

\title{Constrained defined-contribution pension management: A reparameterized neural network approach}

\author{
    Yueman Feng\thanks{\rm Corresponding author. Department of Statistics and Actuarial Science, The University of Hong Kong, Pokfulam, Hong Kong. Email: fengym@connect.hku.hk} \qquad Wenyuan Li\thanks{\rm Department of Statistics and Actuarial Science, The University of Hong Kong, Pokfulam, Hong Kong. Email: wylsaas@hku.hk}
}

\date{\today}

\begin{document}

\maketitle

\begin{abstract}
This paper studies constrained defined contribution (DC) pension management under a two-factor financial market model. The pension participant is allowed to invest in cash, two nominal bonds, an inflation-linked bond, and a stock index. Moreover, they can also purchase term life insurance to manage pre-retirement death risk. Unlike traditional literature, we consider a pension member restricted to a general non-empty, closed, and convex trading constraint set that includes no short-selling, no borrowing, and portfolio-mix constraints as special cases. Such restrictions reflect common limits on pension-account trading. To solve the problem, we construct a family of unconstrained artificial markets, in which the original trading constraints are satisfied by adjusting the drift terms of the risky assets. Using a duality transformation, we derive the Lagrangian dual problem and propose a reparameterized neural network method to compute lower and upper bounds on the primal value function. We calibrate our model with U.S. data using a Kalman filter method, and find that our method produces a smaller duality gap than the benchmark methods. We also compare the constrained results with the unconstrained explicit solution and find that trading constraints substantially change wealth accumulation, portfolio strategies, and insurance demand. This highlights that unconstrained solutions may provide misleading guidance for pension management in practice.

\end{abstract}

% \noindent {\textbf{JEL classification:} C61; D91; G11; G22; H55}\\

\newpage

\section{Introduction}

Defined-contribution (DC) pension plans are playing an increasingly important role in pension systems worldwide. Unlike defined benefit plans, where retirees receive a guaranteed retirement benefit based on salary history, service years, and retirement age, DC pension members make their own investment decisions, and retirement benefits depend on contributions and investment returns. According to the OECD's report, assets in defined contribution plans increased by 11.4\% across OECD countries and other jurisdictions in 2024, compared with the 4.0\% increase in defined benefit plans. Moreover, the proportion of defined benefit plans dropped from 39.7\% in 2014 to 31.9\% in total pension plan assets at the end of 2024 \citep{oecd2025pension}. 

DC pension plans transfer more responsibility from plan sponsors to individual participants. Consequently, the long-horizon investment and insurance decisions become more essential to retirement arrangements. Specifically, during the accumulation period, an individual's account faces a financial risk incorporating interest rate uncertainty, inflation risk, equity risk, and future contribution uncertainty. Meanwhile, participants also face pre-retirement mortality risk, which might leave intended beneficiaries without sufficient financial protection. For the financial risk, extensive literature has already studied DC management in a stochastic environment with different market models and preferences (see \cite{boulier2001optimal}, \cite{battocchio2004optimal}, \cite{han2012optimal}, \cite{yao2013markowitz}, \cite{guan2014optimal}, \cite{chen2015optimal}, \cite{menoncin2017mean}, \cite{tang2018asset},\cite{dong2020optimal}, \cite{wang2021equilibrium}, \cite{chen2023portfolio}, \cite{peng2025optimal}, \cite{gao2025bayesian}, and \cite{ye2026continuous}). For the pre-retirement death risk, the literature can be divided into two groups: the return-of-premiums clause, which returns the contributed premium to the beneficiary during the accumulation phase (see \cite{he2013optimal}, \cite{li2017equilibrium}, \cite{bian2018pre}); the return of account value clause, which returns the DC account value to the beneficiary (see \cite{yao2014asset}, \cite{konicz2015optimal}, \cite{wu2015equilibrium}). More recently, \cite{li2024optimal} introduce life insurance to hedge the pre-retirement death risk and study the optimal DC management under a two-factor model. 

Beyond those two risks, pension management in practice is also subject to trading restrictions and regulatory constraints. Pension accounts are often restricted from short selling, borrowing, and leveraging. Ignoring these constraints can lead to aggressive leverage when maximizing retirement benefits, which can deviate from the optimal path in practice and mislead individuals and pension managers. Compared with the classical unconstrained problem, the constrained problem generally does not have a closed-form solution. Portfolio constraints hinder direct use of martingale methods because the associated dual pricing kernels need not be unique. Existing constrained portfolio optimization literature provides alternative ways of addressing the problem by constructing fictitious markets and using duality methods. \cite{karatzas1991martingale} construct a fictitious market by using additional stocks to complete the market, then by adjusting the auxiliary stock process to make it least favorable in the fictitious market so that the optimal trading strategy is the same as the one under the real market. \cite{cvitanic1992convex} extend this idea to a general nonempty, closed, and convex constraint set by considering a family of fictitious markets in which trading is unconstrained and deriving dual bounds for the original problem. \cite{he1993labor} incorporate labor income into a borrowing constraint problem, and \cite{cuoco1997optimal} extends the problem further to the case with stochastic income and more general constraints. 

In actuarial and pension applications, \cite{zeng2016optimal} study optimal life insurance strategy with no-borrowing constraints. \cite{dong2019optimal} analyze defined contribution pension management under short-selling constraints and portfolio insurance. \cite{li2019data} show how a neural network can learn a DC asset-allocation policy directly from simulated or historical return paths. \cite{hambel2022solving} use artificial insurance markets to solve life cycle problems with biometric risk. \cite{yao2014asset} study the DC management under the S-shaped utility with a path-dependent reference point. They rigorously show the equivalence between the primal and concavified problems under no short-selling constraint. However, existing constrained pension studies usually focus on simpler financial markets, specific types of constraints, or settings without jointly modeled life insurance. There remain research gaps in general trading constraints and a more advanced market model capturing stochastic interest rates and inflation risk.

This paper fills this gap by studying a constrained defined contribution pension management problem with stochastic interest rate and inflation risk. Following \cite{koijen2011optimal}, we adopt a two-factor model to describe the financial market, which allows the real interest rate, expected inflation, and risk premia to vary over time. A pension member is allowed to invest in cash, two nominal bonds with different maturities, an inflation-linked bond, and the stock index. The individual is also allowed to purchase term life insurance before retirement to hedge pre-retirement death risk. Following \cite{li2024constrained}, we restrict the trading strategy to a general non-empty, closed, and convex set. This framework includes no short selling, no borrowing, nontradable assets, and portfolio-mix restrictions as special cases. Inspired by \cite{cuoco1997optimal}, we construct a family of artificial markets by adding drift compensations to the asset processes. By manipulating these compensated terms, we alter the products' attractiveness, inducing the individual to follow the trading rules in the real market. Under trading constraints, the relevant pricing kernels are not uniquely determined, complicating the direct use of martingale methods. We thus adopt a Lagrangian dual control method to find tight lower and upper bounds for the primal problem. For the upper bound, we first derive a family of static budget constraints from the individual's wealth process. Then, we obtain the upper bound for the primal problem by the Lagrangian transform. For the lower bound, we induce a trading strategy from the upper bound, which serves as a candidate strategy for computing the lower bound. 

To reduce variance from high dimensionality, we propose a reparameterized neural network approach. Specifically, we parametrize transformed compensated drift terms directly and recover the original one through a minimal nonnegative shift. After obtaining the dual optimizer, we use classical nested Monte Carlo (bootstrap) \citep[see, e.g.,][]{sun2011efficient} to provide a direct, consistent, and low-bias estimate for the conditional moments in the candidate strategy. Our numerical results compare bootstrap with deep regression and show comparable performance, while bootstrap requires no regression network design or hyperparameter tuning. Moreover, simply truncating the candidate strategy doesn't satisfy the no-borrowing constraints in high dimensions. We therefore propose a Hamiltonian quadratic program to impose the original portfolio constraints on the candidate strategy. Finally, we compare the constrained numerical solution with the unconstrained closed-form benchmark to assess the practical importance of the trading constraints.

The closest numerical benchmark to our method is the simulation of artificial market strategies (SAMS) of \cite{bick2013solving}, which restricts the compensated drift terms to be affine functions of time. This parametrization is computationally convenient but searches only a small subset of the admissible control set. Another close benchmark is the deep pricing kernel approach proposed by \cite{li2024constrained}, which uses a neural network to minimize the upper bound and deep regression to recover a lower-bound policy. It works well in low dimensions but performs poorly in a high-dimensional factor model. Third, the direct neural network parametrization also expands the search space (see \cite{li2019data}), yet it produces an estimate outside our tight dual bounds.

This paper makes three contributions. First, we formulate a DC pension management problem in a two-factor financial market under a general family of convex portfolio constraints and characterize feasible control through the artificial market. Second, we develop a reparameterized neural network method to compute a tight upper bound for a constrained high-dimensional control problem. For the lower bound, we combine nested Monte Carlo estimation of conditional moments with a Hamiltonian quadratic program to recover a feasible candidate strategy under the original portfolio constraints. In both risk-aversion cases, our method reduces the relative duality gap by more than a factor of ten compared with the SAMS and deep pricing kernel benchmarks. We also compare our algorithm with the direct neural network approach of \cite{li2019data}. We find that it is a suboptimal approximation whose utility estimate is lower than our lower bound. Moreover, for $\gamma=1.5$, it produces an optimal insurance strategy that increases monotonically with time, inconsistent with the hump shape widely reported in the literature.  Third, we compare the resulting constrained policies with the unconstrained closed-form benchmark and show that the trading constraints materially change wealth accumulation, asset holdings, and mortality-contingent positions.

The rest of the paper is organized as follows. Section \ref{Economic_setting} introduces the financial market, mortality, wealth process, preference, and portfolio constraint set. Sections \ref{sec:artificial_market} and \ref{sec:primal_dual} develop the artificial market, static budget constraint, primal problem, and dual problem. Section \ref{sec:numerical_analysis} reports the model calibration details, compares the upper and lower bound algorithms, analyzes the optimal solutions, and compares the constrained solution with the unconstrained benchmark. Section \ref{Conclusion} concludes. All proofs are delegated to the appendices.

\section{Economic setting}\label{Economic_setting}
\subsection{Financial market}
We adopt the two-factor financial market of \cite{koijen2011optimal}, in which the real short rate, expected inflation, and risk premia vary over time. Let $(\Omega,\mathcal{F},\mathbb{P})$ be a complete filtered probability space with filtration $\mathbb{F}:=\{\mathcal{F}_t\}_{t\in[0,T]}$. Let the financial risk be described by $Z_t$, an $\mathbb{F}$-adapted four-dimensional vector of independent Brownian motions.

We use a state vector $X_t=(X_{1,t},X_{2,t})^{\top}$ to describe the real short rate and the expected inflation in the financial market:
\begin{equation*}
	r_t=\delta_r+X_{1,t}=\delta_r+e_1^{\top}X_t,\qquad
	\pi_t^e=\delta_{\pi^e}+X_{2,t}=\delta_{\pi^e}+e_2^{\top}X_t,
\end{equation*}
where $\delta_r>0$, $\delta_{\pi^e}>0$, and $e_i$ denotes the $i$-th unit vector in $\mathbb{R}^2$. The two factors follow the Ornstein-Uhlenbeck process
\begin{equation*}\label{eq:X_process}
	dX_t=-K_X X_t\,dt+\Sigma_X\,dZ_t,
\end{equation*}
where $K_X=\operatorname{diag}(\kappa_1,\kappa_2)$ with $\kappa_i>0$, $i=1,2$, and $\Sigma_X=(\sigma_1,\sigma_2)^{\top}$ with $\sigma_i\in\mathbb{R}^4$. This OU process allows both the real short rate and expected inflation to move below zero, which is consistent with episodes of negative real rates and deflation in U.S. history.

The realized inflation process, represented by the consumer price index $\Pi_t$, is
\begin{equation*}\label{eq:realized_inflation}
	\frac{d\Pi_t}{\Pi_t}=\pi_t^e\,dt+\sigma_{\Pi}^{\top}dZ_t,\qquad \Pi_0=1,
\end{equation*}
where $\sigma_{\Pi}\in\mathbb{R}^4$. The equity index $S_t$ evolves according to
\begin{equation*}
	\frac{dS_t}{S_t}=\mu_t\,dt+\sigma_S^{\top}dZ_t,
\end{equation*}
with nominal expected return $\mu_t=R_t+\mu_0+\mu_1^{\top}X_t$. Here $R_t$ is the nominal short rate given in \eqref{eq:nominal_short_rate} below. For identification, the volatility matrix $(\sigma_1,\sigma_2,\sigma_{\Pi},\sigma_S)^{\top}$ is assumed to be lower triangular.

The nominal pricing kernel $\phi_t$ satisfies
\begin{equation*}
	\frac{d\phi_t}{\phi_t}=-R_t\,dt-\Lambda_t^{\top}dZ_t,\qquad \phi_0=1,
\end{equation*}
where the Sharpe ratio $\Lambda_t$ is affine in the state variables:
\begin{equation*}
	\Lambda_t=\Lambda_0+\Lambda_1X_t.
\end{equation*}
Following \cite{koijen2011optimal}, we impose
\begin{equation*}
	\Lambda_0=
	\begin{pmatrix}
		\Lambda_{0(1)}\\
		\Lambda_{0(2)}\\
		0\\
		\Lambda_{0(4)}
	\end{pmatrix},
	\qquad
	\Lambda_1=
	\begin{pmatrix}
		\Lambda_{1(1,1)} & 0\\
		0 & \Lambda_{1(2,2)}\\
		0 & 0\\
		\Lambda_{1(4,1)} & \Lambda_{1(4,2)}
	\end{pmatrix},
\end{equation*}
with $\sigma_S^{\top}\Lambda_0=\mu_0$ and $\sigma_S^{\top}\Lambda_1=\mu_1^{\top}$. The real pricing kernel, defined by $\phi_t^R=\phi_t\Pi_t$, then satisfies
\begin{equation*}
	\frac{d\phi_t^R}{\phi_t^R}
	=-r_t\,dt-(\Lambda_t-\sigma_{\Pi})^{\top}dZ_t
	=-r_t\,dt+\kappa_{0,t}^{\top}dZ_t,\qquad
	\kappa_{0,t}=-(\Lambda_t-\sigma_{\Pi}),\quad \phi_0^R=1.
\end{equation*}
This implies the instantaneous nominal short rate
\begin{equation}\label{eq:nominal_short_rate}
	R_t=\delta_R+(\iota_2^{\top}-\sigma_{\Pi}^{\top}\Lambda_1)X_t,
\end{equation}
where $\delta_R=\delta_r+\delta_{\pi^e}-\sigma_{\Pi}^{\top}\Lambda_0$ and $\iota_2=(1,1)^{\top}$.

Following the standard derivation in the literature \citep[e.g.][]{duffie1996yield}, we obtain the following prices of nominal and inflation-linked bonds. The time-$t$ price of a nominal bond maturing at $s$ is
\begin{equation*}
	P(X_t,t,s)=\exp\{A_0(s-t)+[A_1(s-t)]^{\top}X_t\},
\end{equation*}
where, for $\tau=s-t$, $A_0(\tau)$ and $A_1(\tau)$ solve
\begin{align*}
	\frac{\partial A_0(\tau)}{\partial \tau}
	&=\frac{1}{2}A_1(\tau)^{\top}\Sigma_X\Sigma_X^{\top}A_1(\tau)
	-A_1(\tau)^{\top}\Sigma_X\Lambda_0-\delta_R,
	& A_0(0)&=0, \\
	\frac{\partial A_1(\tau)}{\partial \tau}
	&=-(K_X^{\top}+\Lambda_1^{\top}\Sigma_X^{\top})A_1(\tau)-\iota_2+\Lambda_1^{\top}\sigma_{\Pi},
	& A_1(0)&=0. 
\end{align*}
The corresponding dynamics of $P(X_t,t,s)$ satisfy
\begin{equation*}
	\frac{dP(X_t,t,s)}{P(X_t,t,s)}
	=\{R_t+[A_1(s-t)]^{\top}\Sigma_X\Lambda_t\}\,dt
	+[A_1(s-t)]^{\top}\Sigma_X\,dZ_t.
\end{equation*}

The time-$t$ real price of an inflation-linked bond with maturity $s$ is then
\begin{equation*}
	P^R(X_t,t,s)=\exp\{A_0^R(s-t)+[A_1^R(s-t)]^{\top}X_t\},
\end{equation*}
where $A_0^R(\tau)$ and $A_1^R(\tau)$ satisfy
\begin{align*}
	\frac{\partial A_0^R(\tau)}{\partial \tau}
	&=\frac{1}{2}A_1^R(\tau)^{\top}\Sigma_X\Sigma_X^{\top}A_1^R(\tau)
	-A_1^R(\tau)^{\top}\Sigma_X(\Lambda_0-\sigma_{\Pi})-\delta_r,
	& A_0^R(0)&=0,\\
	\frac{\partial A_1^R(\tau)}{\partial \tau}
	&=-(K_X^{\top}+\Lambda_1^{\top}\Sigma_X^{\top})A_1^R(\tau)-e_1,
	& A_1^R(0)&=0.
\end{align*}
The corresponding nominal price $\Pi_tP^R(X_t,t,s)$ of the inflation-linked bond has dynamics
\begin{equation*}
	\frac{d(\Pi_tP^R(X_t,t,s))}{\Pi_tP^R(X_t,t,s)}
	=\{R_t+[A_1^R(s-t)]^{\top}\Sigma_X\Lambda_t+\sigma_{\Pi}^{\top}\Lambda_t\}\,dt
	+\{[A_1^R(s-t)]^{\top}\Sigma_X+\sigma_{\Pi}^{\top}\}\,dZ_t.
\end{equation*}

\subsection{Mortality}
We next introduce mortality risk. Let $T_x$ denote the future lifetime of an individual aged $x$, which is non-negative and is assumed to be independent of the financial market filtration $\mathbb{F}$. For $t\geq0$, define the survival and death probabilities by
\begin{equation*}
	{}_{t}p_x=\mathbb{P}(T_x>t),\qquad
	{}_{t}q_x=\mathbb{P}(T_x\leq t)=1-{}_{t}p_x.
\end{equation*}
We then have $\lim_{t\to\infty}{}_{t}p_x=0$ and $\lim_{t\to\infty}{}_{t}q_x=1$. Here, ${}_{t}p_x$ is the probability that the individual survives from age $x$ to age $x+t$, while ${}_{t}q_x$ is the probability of death before age $x+t$.

The instantaneous force of mortality, or hazard rate, is denoted by $\lambda_{x+t}$ and satisfies
\begin{equation*}
	\lambda_{x+t}
	=\frac{1}{{}_{t}p_x}\frac{d}{dt}{}_{t}q_x
	=-\frac{1}{{}_{t}p_x}\frac{d}{dt}{}_{t}p_x.
\end{equation*}
As a result, the survival and death probabilities can be written in terms of the force of mortality:
\begin{equation*}
	{}_{t}p_x=\exp\left\{-\int_0^t \lambda_{x+s}\,ds\right\},
	\qquad
	{}_{t}q_x=\int_0^t{}_{s}p_x\lambda_{x+s}\,ds.
\end{equation*}
Hence the density of $T_x$ is $f_{T_x}(t)={}_{t}p_x\lambda_{x+t}$ for $t>0$.

\subsection{Wealth process}\label{subsec:wealth_process}
The individual joins the DC pension plan at age $x$ and retires at time $T$. During the accumulation period $0\leq t<T\wedge T_x$, the individual contributes continuously to the pension account. The real contribution rate $C_t^R$ is assumed to be deterministic and satisfies
\begin{equation*}\label{eq:real_contribution}
	\frac{dC_t^R}{C_t^R}=g_t^R\,dt,\qquad 0\leq t<T\wedge T_x,
\end{equation*}
where $g_t^R$ is the growth rate of the real contribution rate. Let $C_t$ denote the corresponding nominal contribution rate, so that $C_t^R=C_t/\Pi_t$.

The investment opportunity set consists of a cash account, two nominal bonds, one inflation-linked bond, and the equity index. We use the rolling bond strategy for the bond positions as in \cite{boulier2001optimal}. At each date, the proceeds from a maturing position are rolled into a newly issued bond with the same remaining maturity. Let $\theta_t=(\theta_{1,t},\theta_{2,t},\theta_{3,t},\theta_{4,t})^{\top}$ be the nominal amounts invested in the nominal bond with maturity $T_1$, the nominal bond with maturity $T_2$, the inflation-linked bond with maturity $T_3$, and the equity index, respectively. The amount invested in the cash account is denoted by $\alpha_t$, and hence $W_t=\alpha_t+\iota_4^{\top} \theta_t$, where $\iota_4=(1,1,1,1)^{\top}$.

The volatility matrix of the four non-cash assets is
\begin{equation*}
	\Sigma=
	\begin{pmatrix}
		[A_1(T_1)]^{\top}\Sigma_X\\
		[A_1(T_2)]^{\top}\Sigma_X\\
		[A_1^R(T_3)]^{\top}\Sigma_X+\sigma_{\Pi}^{\top}\\
		\sigma_S^{\top}
	\end{pmatrix}.
\end{equation*}
While alive and before retirement, the individual may also purchase term life insurance with premium rate denoted by $I_t$. The allowance for life insurance in a 401(k) pension fund can be found in \cite{TreasReg1401}. If the individual dies at time $T_x=t$ before retirement, the designated beneficiary receives a death benefit equal to the face value of the life insurance, $I_t/\lambda_{x+t}$, together with the account balance. We assume insurance contracts are continuously tradable. Let $D_t$ denote the cumulative free disposal, namely the amount of money the individual chooses not to reinvest up to time $t$. The nominal wealth process then follows
\begin{equation*}
dW_t=\theta_t^{\top}\Sigma\Lambda_t\,dt+W_tR_t\,dt+C_t\,dt+\theta_t^{\top}\Sigma\,dZ_t-I_t\,dt-dD_t,\qquad 0\leq t<T\wedge T_x.
\end{equation*}

We can then derive the real wealth dynamics $W_t^R=W_t/\Pi_t$:
\begin{equation}\label{eq:real_wealth}
	dW_t^R
	=\left\{r_tW_t^R+\left[(\theta_t^R)^{\top}\Sigma-W_t^R\sigma_{\Pi}^{\top}\right](\Lambda_t-\sigma_{\Pi})\right\}dt
	+\left[(\theta_t^R)^{\top}\Sigma-W_t^R\sigma_{\Pi}^{\top}\right]dZ_t
	+C_t^R\,dt-I_t^R\,dt-dD_t^R,
\end{equation}
where $W_0^R=w^R$, $\theta_t^R=\theta_t/\Pi_t$, $I_t^R=I_t/\Pi_t$, and $dD_t^R=\Pi_t^{-1}dD_t$. Here, we assume the wealth process is lower-bounded by the negative expected future contribution 
\begin{equation*}
    W^R_t \geq -\int_{t}^T {_{s-t}p_{x+t}}  P^R(X_t,t,s) C^R_s ds. 
\end{equation*}
It is easy to verify that this lower bound is finite almost surely by the continuity of deterministic functions $_{s-t}p_{x+t}$, $A^R_0$, $A^R_1$, and $C^R_s$ over $[t,T]$.
\begin{eqnarray}
    W^R_t &\geq& -\int_{t}^T {_{s-t}p_{x+t}}  P^R(X_t,t,s) C^R_s ds \notag\\ 
    &\geq&-\int_{t}^T \exp(c+c|X_t|)ds \notag\\
    &=& -(T-t)\exp(c+c|X_t|)>-\infty ~~\text{a.s.}, \label{wealth_finite_LB} 
\end{eqnarray}
where $c$ is some real constant. The last inequality follows because the OU process $X_t$ is finite a.s. This finite lower bound (credit limit) avoids arbitrage opportunities, such as the doubling strategy in \cite{harrison1979martingales}. Moreover, we also assume $W^R_T \geq 0$, meaning the individual has no deficit at retirement time. Lastly, the real death benefit $M_t^R$ is
\begin{equation} \label{real_death_benefit}
	M_t^R=W_t^R+\frac{I_t^R}{\lambda_{x+t}},\qquad T_x=t<T.
\end{equation}

\subsection{Preference and feasibility}\label{subsec:preference_feasibility}
The individual chooses investment and insurance strategies to maximize expected utility from the payoff received at death or at retirement, whichever occurs first. 
\begin{equation*}
	\sup_{(\alpha^R,\theta^R)\in A,\,I^R}
	E\left[U(M_{T_x}^R)\1_{\{T_x<T\}}+U(W_T^R)\1_{\{T_x\geq T\}}\right],
\end{equation*}
where $A\subset\mathbb{R}^{5}$ is the portfolio constraint set specified later. We assume that the utility function satisfies the following condition.

\begin{definition}[Utility function]\label{def:utility}
The utility function $U:(0,\infty)\rightarrow\mathbb{R}$ is increasing, strictly concave, and continuously differentiable.
\end{definition}

Since $T_x$ is independent of the financial market filtration, this objective can be written on the fixed horizon $[0,T]$ as
\begin{equation}\label{eq:objective}
	\sup_{(\alpha^R,\theta^R)\in A,\,I^R}
	E\left[
		\int_0^T {}_{t}p_x\lambda_{x+t}U(M_t^R)\,dt
		+{}_{T}p_xU(W_T^R)
	\right].
\end{equation}
Here, we don't add a constraint on $I^R_t$. Because the cash flows of life insurance and annuities are opposite, $I^R_t<0$ measures individual demand for the annuity(see \cite{fischer1973life} and \cite{li2025optimal}). Adding non-negative constraints on $I^R_t$ ignores the individual demand for the annuity in a DC plan. 

We next define the set of bequest-terminal plans. Let $Q_0$ be the reference risk-neutral measure associated with the real pricing kernel, under which $dZ_{0,t}=dZ_t-\kappa_{0,t}\,dt$ is a Brownian motion. Define
\begin{equation}\label{eq:bequest_terminal_set}
	G:=\left\{(M^R,W_T^R):
	E^{Q_0}\left[\int_0^T |M_t^R|\,dt+|W_T^R|\right]<\infty
	\right\}.
\end{equation}
Let $G_+$ denote the non-negative orthant of $G$, so $M_t^R\geq0$ for $0\leq t\leq T$ and $W_T^R\geq0$. We write $G_+^*$ for the set of plans $(M^R,W_T^R)\in G_+$ satisfying
\begin{align*}
	&\min\left\{
	E\left[\int_0^T {}_{t}p_x\lambda_{x+t}U(M_t^R)^+\,dt\right],
	E\left[\int_0^T {}_{t}p_x\lambda_{x+t}U(M_t^R)^-\,dt\right]
	\right\}<\infty,\\
	&\min\left\{
	E\left[{}_{T}p_xU(W_T^R)^+\right],
	E\left[{}_{T}p_xU(W_T^R)^-\right]
	\right\}<\infty,
\end{align*}
where $y^+=\max\{y,0\}$ and $y^-=\max\{-y,0\}$. These conditions ensure that the expected utility in \eqref{eq:objective} is well-defined in $(-\infty,+\infty)$.

Let $\Theta$ be the set of progressively measurable controls $(\alpha^R,\theta^R,I^R)$ satisfying $W_t^R=\alpha_t^R+\iota_4^{\top}\theta_t^R$ and
\begin{equation*}
	\int_0^T\left(
		|r_tW_t^R|
		+\left|\left[(\theta_t^R)^{\top}\Sigma-W_t^R\sigma_{\Pi}^{\top}\right](\Lambda_t-\sigma_{\Pi})\right|
		+\left\|(\theta_t^R)^{\top}\Sigma-W_t^R\sigma_{\Pi}^{\top}\right\|^2
		+|I_t^R|
	\right)dt<\infty.
\end{equation*}
It remains to specify what it means for a plan to be financed. Given the market parameters $\mathcal{P}$, a plan $(M^R,W_T^R)\in G_+^*$ is feasible if there exist $(\alpha^R,\theta^R,I^R)\in\Theta$ and a non-negative, non-decreasing free-disposal process $D^R$ such that the real wealth process \eqref{eq:real_wealth} starts from $W_0^R=w^R_0$, satisfies $W_T^R\geq0$, and finances the death benefit through $M_t^R=W_t^R+I_t^R/\lambda_{x+t},\ 0\leq t<T$. The plan is $A$-feasible if, in addition, $(\alpha_t^R,\theta_t^R)\in A$ for all $t\in[0,T]$. We denote by $\mathcal{B}(\mathcal{P},A)$ the set of $A$-feasible bequest-terminal plans under the market parameters $\mathcal{P}$.

\subsection{Portfolio constraint set} \label{sec_portfolio_constraint_set}
We assume that the individual's trading decisions $(\alpha^R,\theta^R)$ are restricted by a portfolio constraint set $A$. Following the convex-set formulation of \cite{cuoco1997optimal}, the pair $(\alpha^R,\theta^R)$ take values in a non-empty, closed, convex set $A\subset\mathbb{R}^{5}$. For $u=(u_0,u_-)\in\mathbb{R}\times\mathbb{R}^4$, define the support function of $-A$ by
\begin{equation}\label{eq:support_function}
	\delta(u)=\sup_{(\alpha^R,\theta^R)\in A}
	-\left(\alpha^R u_0+(\theta^R)^{\top}u_-\right).
\end{equation}
To identify where this support function is finite, we define its effective domain as
\begin{equation}\label{eq:effective_domain}
	\widetilde{A}:=\{u\in\mathbb{R}^{5}:\delta(u)<\infty\}.
\end{equation}
By standard results in convex analysis, the function $\delta$ is positively homogeneous, lower semicontinuous, proper, and convex on $\mathbb{R}^{5}$, and $\widetilde{A}$ is a closed convex cone. We assume the supporting function satisfies the assumption below
\begin{assumption}\label{support_function_assumption}
  The supporting function $\delta$ is upper semi-continuous and bounded above in $\widetilde{A}$. Futhermore, $u_0\geq 0$ for any $u \in \widetilde{A}$. 
\end{assumption}
We can easily obtain $u_0\geq 0$ for any $u \in \widetilde{A}$, as one can invest a large amount of money only in the cash account $(\alpha,0) \in A$, and a negative $u_0$ would send the supporting function to infinity. Furthermore, because $\delta$ is positively homogeneous and $\widetilde{A}i$ is a cone, $\delta$ is bounded above on $\widetilde{A}$ if and only if $\delta$ is non-positive on $\widetilde{A}$.

This formulation can represent non-tradable assets, short-sale constraints, buying constraints, portfolio-mix constraints, minimum capital requirements, and combinations of these restrictions; see \cite{cvitanic1992convex,cuoco1997optimal}. In this paper, we focus on the portfolio-mix constraint with $D=[0,1]$. This portfolio-mixed constraint is a ``long-only constraint'' that comes from \cite{IRC4975}, which states that ``lending of money or other extension of credit is prohibited in a 401(k) pension fund.'' Under this constraint, both the cash account position and the four non-cash asset positions are non-negative:
\begin{equation}\label{eq:portfolio_constraint_set}
	A=\left\{(\alpha^R,\theta^R)\in\mathbb{R}^{5}:
	\alpha^R\geq0,\quad \theta^R\geq0
	\right\},
\end{equation}
where the inequality for $\theta^R$ is componentwise. The corresponding effective domain is
\begin{equation}\label{eq:effective_domain_specific}
	\widetilde{A}=
	\left\{(u_0,u_-)\in\mathbb{R}^{5}:
	u_0\geq0,\quad u_-\geq0
	\right\}.
\end{equation}
 For $u\in\widetilde{A}$, the supremum in \eqref{eq:support_function} is attained at $\alpha^R=0$ and $\theta^R=0$, so $\delta(u)=0$. If at least one component of $u$ is negative, then $\delta(u)=+\infty$. Hence the effective domain condition \eqref{eq:effective_domain} will require the relevant vector $u$ to lie in $\widetilde{A}$. 

\section{Artificial market and static budget constraint}\label{sec:artificial_market}
Our main challenge is to handle the portfolio optimization problem under the portfolio constraint set $A$. Following \cite{cuoco1997optimal}, rather than directly restricting trading decisions, we construct an unconstrained artificial market by introducing a compensated process $v_t$ to adjust the drift of asset prices. With an appropriate adjustment, any rational individual seeks to maximize expected utility will choose to trade in an $A$-feasible way. 

Given the constraint set $A$, let $\mathcal{N}$ be the set of progressively measurable processes $v=(v_0,v_-)$ 
\begin{equation*}
	E\left[\int_0^T |v_t|^2\,dt\right]<\infty.
\end{equation*}
For each $v\in\mathcal{N}$, the processes
\begin{align}
	\beta_{v,t}
	&=\exp\left\{-\int_0^t (r_s+v_{0,s})\,ds\right\}, \label{eq:beta_v}\\
	\kappa_{v,t}
	&=-(\Lambda_t+v_{-,t}-\sigma_{\Pi}), \label{eq:kappa_v}\\
	\xi_{v,t}
	&=\exp\left\{\int_0^t \kappa_{v,s}^{\top}dZ_s
	-\frac{1}{2}\int_0^t|\kappa_{v,s}|^2\,ds\right\}, \label{eq:xi_v}\\
	\pi_{v,t}
	&=\beta_{v,t}\xi_{v,t}, \label{eq:pi_v}\\
	dZ_{v,t}
	&=dZ_t-\kappa_{v,t}\,dt. \label{eq:Zv}
\end{align}
define the artificial market $\mathcal{M}_v$, where $\xi_v$ is a strictly positive local martingale. To ensure that this fictitious unconstrained market is arbitrage-free, $\xi_v$ must be a true martingale. Thus, we define $\mathcal{N}^*$ as the subset of $\mathcal{N}$ for which $\xi_v$ is a true martingale. Novikov's condition, together with the cone structure of $\widetilde A$, further guarantees that $\mathcal{N}^*$ is non-empty.

For any $v\in\mathcal{N}^*$, $\pi_{v,t}$ is the unique state-price density in the artificial market $\mathcal{M}_v$. By adjusting the drift terms through $v=(v_0,v_-)$, the risky assets can become more or less attractive relative to the cash account. We are thus able to induce the individual to trade in an $A$-feasible way. Theorem~\ref{theorem:static_budget_constraint} characterizes feasibility through the family of static budget constraints for all admissible artificial markets.
\begin{theorem}\label{theorem:static_budget_constraint}
A bequest-terminal plan $(M^R,W_T^R)\in G_+^*$ is $A$-feasible if and only if
\begin{align}
	&E^{Q_v}\left[
		{_{T}p_x}\beta_{v,T} W_T^R
		+\int_0^T {_{t}p_x}\beta_{v,t}\lambda_{x+t}M_t^R\,dt
	\right] \leq
	w^R_0+
	E^{Q_v}\left[
		\int_0^T {_{t}p_x}\beta_{v,t}
		\left(C_t^R+\delta(\widetilde v_t)\right)dt
	\right],
	\,\, \forall v\in\mathcal{N}^*. \label{eq:static_budget_constraint}
\end{align}
where
\begin{equation}\label{eq:tilde_v_mapping}
	\widetilde{v}_{0,t}=v_{0,t}-\sigma_{\Pi}^{\top}v_{-,t},\qquad
	\widetilde{v}_{-,t}=\Sigma v_{-,t}+\widetilde{v}_{0,t}\iota_4.
\end{equation}
\end{theorem}

\begin{remark}\label{remark:v transformation}
To complete the supporting function \eqref{eq:support_function}, we need a transform of $v$ to $\widetilde{v}$ under the two-factor model. The transformation $v=(v_0,v_-)\mapsto\widetilde v=(\widetilde v_0,\widetilde v_-)$ in \eqref{eq:tilde_v_mapping} is linear and preserves square integrability. Specifically,
\begin{equation*}
	v=(v_0,v_-)\in L^2([0,T]\times\Omega;\mathbb{R}^5)
	\quad\Longrightarrow\quad
	\widetilde v=(\widetilde v_0,\widetilde v_-)\in L^2([0,T]\times\Omega;\mathbb{R}^5).
\end{equation*}
Consequently, we restrict $\widetilde v \in \widetilde A$ rather than the original $v$. For brevity, $\widetilde{v}\in\mathcal{N}^*$ means $v\in\mathcal{N}^*$ throughout the paper.
\end{remark}

% This yields the following direct corollary.
% \begin{corollary}\label{corollary:no_free_disposal}
% Suppose there exists $v^*\in\mathcal{N}^*$ such that, for every $v\in\mathcal{N}^*$,
% \begin{align}
% 	&E^{Q_v}\left[
% 		{_{T}p_x}\beta_{v,T} W_T^R
% 		+\int_0^T {_{t}p_x}\beta_{v,t}\lambda_{x+t}M_t^R\,dt
% 		-\int_0^T {_{t}p_x}\beta_{v,t}
% 		\left(C_t^R+\delta(\widetilde v_t)\right)dt
% 	\right] \notag\\
% 	&\qquad\leq
% 	E^{Q_{v^*}}\left[
% 		{_{T}p_x}\beta_{v^*,T} W_T^R
% 		+\int_0^T {_{t}p_x}\beta_{v^*,t}\lambda_{x+t}M_t^R\,dt
% 		-\int_0^T {_{t}p_x}\beta_{v^*,t}
% 		\left(C_t^R+\delta(\widetilde v_t^*)\right)dt
% 	\right]
% 	=w_0. \label{eq:corollary_condition}
% \end{align}
% Then $(M^R,W_T^R)$ is feasible. The financed wealth process is
% \begin{align}
% 	W_{v^*,t}^R
% 	=E_t^{Q_{v^*}}\bigg[
% 	&\int_t^T
% 	\exp\left\{-\int_t^s(r_u+v_{0,u}^*+\lambda_{x+u})\,du\right\}
% 	\left(\lambda_{x+s}M_s^R-C_s^R-\delta(\widetilde v_s^*)\right)ds \notag\\
% 	&+\exp\left\{-\int_t^T(r_s+v_{0,s}^*+\lambda_{x+s})\,ds\right\}
% 	W_T^R
% 	\bigg], \label{eq:optimal_wealth_corollary}
% \end{align}
% and the optimal free-disposal process satisfies $D_t^{R,*}\equiv0$.
% \end{corollary}

\section{Primal problem and dual problem}\label{sec:primal_dual}
Theorem \ref{theorem:static_budget_constraint} allows us to formulate the primal problem as
\manualtaglabel{eq:primal_problem}{P}
\begin{align}
	 &\qquad \qquad \qquad  \qquad \qquad \qquad  \sup_{(M^R,W_T^R)\in G_+^*} J(M^R,W_T^R) \tag*{(P)}\label{primal_problem}\\
	\text{s.t.}\quad &
	E^{Q_v}\left[
		{_{T}p_x}\beta_{v,T} W_T^R
		+\int_0^T {_{t}p_x}\beta_{v,t}\lambda_{x+t}M_t^R\,dt
	\right] \leq
	w^R_0+
	E^{Q_v}\left[
		\int_0^T {_{t}p_x}\beta_{v,t}
		\left(C_t^R+\delta(\widetilde v_t)\right)dt
	\right],\, \forall v\in\mathcal{N}^*. \notag
\end{align}
where
\begin{equation}\label{eq:primal_objective}
	J(M^R,W_T^R)
	=
	E\left[
		\int_0^T {}_{t}p_x\lambda_{x+t}U(M_t^R)\,dt
		+{}_{T}p_xU(W_T^R)
	\right].
\end{equation}
Because $0\in\mathcal{N}^*$, problem \eqref{eq:primal_problem} is a convex optimization problem over a closed and bounded subset of $L^1(\lambda \otimes Q_0)$, where $\lambda$ denotes Lebesgue measure on $[0,T]$. However, $L^1$ is not reflexive, and closed bounded subsets of $L^1$ need not be compact. To handle this, we use the Lagrangian dual control method to solve the optimization problem.

Because $\{\pi_v: v\in \mathcal{N}^*\}$ is convex, this suggests the existence of a pricing kernel $\pi_{v^*}$ and a Lagrange multiplier $\psi^*>0$ such that $(M^*, W^*_T,\psi^*,v^*)$ is a saddle point of the Lagrangian:
\begin{align}
	\mathcal{L}(M^R,W_T^R,\psi,v)
	&=E\left[
		\int_0^T {}_{t}p_x\lambda_{x+t}U(M_t^R)\,dt
		+{}_{T}p_xU(W_T^R)
	\right] \notag\\
	&+\psi\bigg\{
		w^R_0-E\left[\int_0^T {_{t}p_x}\pi_{v,t}\left(\lambda_{x+t}M_t^R-C_t^R-\delta(\widetilde v_t)\right)dt+{_{T}p_x}\pi_{v,T}W_T^R\right]
	\bigg\}. \label{eq:lagrangian}
\end{align}
It is straightforward to see maximizing $\mathcal{L}(M^R,W_T^R,\psi,v)$ over $(M^R,W_T^R)$ and minimizing over $(\psi,v)$ yields an upper bound for the primal problem. Specifically, 
\begin{align}
    J(M^R,W_T^R)&:=E\left[
		\int_0^T {_{t}p_x}\lambda_{x+t}U(M_t^R)\,dt
		+{}_{T}p_xU(W_T^R)\right]\notag\\
    &=E\left[
		\int_0^T {_{t}p_x}\lambda_{x+t}[U(M_t^R)-\psi \pi_{v,t} M^R_t]\,dt 
		+{_{T}p_x}[U(W_T^R)-\psi\pi_{v,T}W^R_T]\right]\notag\\
    &+\psi E\left[\int_0^T {_tp_x}\lambda_{x+t} \pi_{v,t}M^R_t dt + {_Tp_x}\pi_{v,T}W^R_T\right]\notag\\
    &\leq E\left[
		\int_0^T {_{t}p_x}\lambda_{x+t}\sup \limits_{M^R_t>0}[U(M_t^R)-\psi \pi_{v,t} M^R_t]\,dt 
		+{_{T}p_x}\sup \limits_{W^R_T>0}[U(W_T^R)-\psi\pi_{v,T}W^R_T]\right]\notag\\
    &+\psi \left\{ w^R_0 + E\left[\int_0^T {_tp_x}\pi_{v,t}[C^R_t+\delta(\widetilde v_t)]dt\right] \right\}\notag\\
    &:=\sup \limits_{M^R>0,\, W^T_R>0}\mathcal{L}(M^R,W_T^R,\psi,v),\label{lag_upperbound}
\end{align} 
where the first inequality comes from taking the supremum with $(M,W_T)$ and the static budget constraint \eqref{eq:static_budget_constraint}. Finally, taking the supremum on the left and the infimum on the right of \eqref{lag_upperbound}, we obtain
\begin{equation*}
    \sup \limits_{M^R>0, \, W^R_T>0} J(M^R,W_T^R) \leq \inf_{(\psi,v)\in(0,\infty)\times\mathcal{N}^*}\sup \limits_{M^R>0,\, W^T_R>0}\mathcal{L}(M^R,W_T^R,\psi,v)
\end{equation*}
Define $\widetilde{J}(\psi,v)$ as the Lagrangian $\mathcal{L}$ maximizing with $(M^R, W^R_T)$, then we derive the dual problem
\manualtaglabel{eq:dual_problem}{D}
\begin{equation*}\tag*{(D)}
	\inf_{(\psi,v)\in(0,\infty)\times\mathcal{N}^*}\widetilde J(\psi,v),
\end{equation*}
where
\begin{align}
	\widetilde J(\psi,v)
	=
	E\bigg[
	&\int_0^T {}_{t}p_x\lambda_{x+t}\widetilde U(\psi\pi_{v,t})\,dt
	+{}_{T}p_x\widetilde U(\psi\pi_{v,T}) +\psi\left(
		w^R_0+\int_0^T {_{t}p_x}\pi_{v,t}\left(C_t^R+\delta(\widetilde v_t)\right)dt
	\right)
	\bigg], \label{eq:dual_objective}
\end{align}
and the dual utility is defined by
\begin{equation}\label{eq:dual_utility}
	\widetilde U(z)=\sup_{x>0}\{U(x)-zx\},\qquad z>0.
\end{equation}
\begin{remark}[Properties of the dual utility]\label{lemma:dual_utility}
For a utility function $U$ satisfying the Inada conditions
\begin{equation}\label{eq:inada}
	U'(0+)=\infty,\qquad U'(\infty)=0,
\end{equation}
the dual utility $\widetilde U$ defined in \eqref{eq:dual_utility} has the following properties:
\begin{enumerate}
	\item $\widetilde U:(0,\infty)\to\mathbb{R}$ is strictly decreasing and strictly convex.
	\item $\widetilde U(0+)=U(\infty),\qquad, \widetilde U(\infty)=U(0+)$.
	\item Define
	\begin{equation}\label{eq:f_definition}
		f(z)=(U')^{-1}(z),\quad z>0,
	\end{equation}
	then $\widetilde U(z)=U(f(z))-zf(z)$.
	\item $\widetilde U'(z)=-f(z)$. 
\end{enumerate}
\end{remark}

Lastly, fixing the artificial market reduces the dual problem to a one-dimensional optimization over $\psi$. We have an interesting result that the associated optimal wealth process under fixed market $\mathcal{M}_v$ has zero free disposal.
\begin{proposition}\label{prop:fixed_v_dual}
For any fixed $v\in\mathcal{N}^*$, there exists a unique $\psi_v>0$ that minimizes $\widetilde J(\psi,v)$ and satisfies
\begin{equation}\label{eq:psi_foc}
	\frac{\partial \widetilde J(\psi_v,v)}{\partial \psi}=0.
\end{equation}
The associated wealth process is
\begin{equation}
	W_{v,t}^R
	=E_t^{Q_v}\bigg[
	\int_t^T
	{_{s-t}p_{x+t}}\frac{\beta_{v,s}}{\beta_{v,t}}
	\left(\lambda_{x+s}f(\psi_v\pi_{v,s})-C_s^R-\delta(\widetilde v_s)\right)ds +{_{T-t}p_{x+t}}\frac{\beta_{v,T}}{\beta_{v,t}}
	f(\psi_v\pi_{v,T})
	\bigg], \label{eq:wealth_fixed_v}
\end{equation}
and the optimal free-disposal process is zero.
\end{proposition}

\section{Numerical analysis}\label{sec:numerical_analysis}
\subsection{Model calibration and numerical setting}\label{subsec:model_calibration}
The financial market parameters are estimated by a Kalman filter in the state-space representation \citep{harvey1990forecasting} with monthly U.S. data from June 1961 to September 2025. Following \cite{gurkaynak2007us}, we use zero-coupon nominal yields at eight maturities: three months, six months, one year, two years, three years, five years, seven years, and ten years. The realized inflation series is the FRED CPIAUCNS series obtained from \cite{fredCPIAUCNS}, and the equity index is obtained from the CRSP value-weighted NYSE/Amex/Nasdaq return with dividends reinvested. We select the maximum-likelihood estimate with the highest log-likelihood across multiple starting values. The annualized estimates are reported in Table \ref{tab:financial_market_estimate_table}. As in \cite{koijen2011optimal}, the estimate $\kappa_1>\kappa_2$ implies that expected inflation is more persistent than the real short rate. The estimate $\sigma_{2(1)}<0$ implies a negative instantaneous covariance between the real short rate and expected inflation factors. For the equity index process, both $\mu_{1(1)}$ and $\mu_{1(2)}$ are negative, so the equity risk premium declines when either factor rises. Moreover, the unconditional price of risk $\Lambda_0$ is positive for the real short rate and equity and negative for expected inflation, while the conditional factor loadings in $\Lambda_1$ are negative.

\begin{table}[htbp]
	\caption{\textbf{Estimation results for the financial market}}\label{tab:financial_market_estimate_table}
	\smallskip
	\centering
	\begin{tabular}{c r r r r}
		Parameters &Estimate &Standard Error &t-statistics&p-value\\
		\hline
		\multicolumn{5}{l}{Average short rate \& average expected inflation}\\
		$\delta_r$ & 0.01198 & 0.00490 & 2.44474 &0.01450\\
		$\delta_R$ & 0.04982 & 0.01265 & 3.93743 &0.00008\\
		$\delta_{\pi^e}$ & 0.03751 & 0.01284 & 2.92107 &0.00349\\
		\multicolumn{5}{l}{Two-factor process}\\
		$\kappa_1$ & 0.62730 & 0.09605 & 6.53086 &0.00000\\
		$\kappa_2$ & 0.19437 & 0.06297 & 3.08662 &0.00202\\
		$\sigma_{1(1)}$ & 0.01873 & 0.00145 & 12.95685 &0.00000\\
		$\sigma_{2(1)}$ & -0.00757 & 0.00135 & -5.61951 &0.00000\\
		$\sigma_{2(2)}$ & 0.01395 & 0.00034 & 41.18751 &0.00000\\
		\multicolumn{5}{l}{Realized inflation process}\\
		$\sigma_{\Pi(1)}$ & 0.00089 & 0.00047 & 1.91484 &0.05551\\
		$\sigma_{\Pi(2)}$ & 0.00212 & 0.00060 & 3.50335 &0.00046\\
		$\sigma_{\Pi(3)}$ & 0.01351 & 0.00034 & 39.51051 &0.00000\\
		\multicolumn{5}{l}{Equity index process}\\
		$\mu_0$ & 0.04663 & 0.03163 & 1.47434 &0.14039\\
		$\mu_{1(1)}$ & -1.97907 & 0.98530 & -2.00860 &0.04458\\
		$\mu_{1(2)}$ & -1.41777 & 0.59671 & -2.37600 &0.01750\\
		$\sigma_{S(1)}$ & -0.01858 & 0.00473 & -3.92803 &0.00009\\
		$\sigma_{S(2)}$ & -0.01790 & 0.00588 & -3.04254 &0.00235\\
		$\sigma_{S(3)}$ & -0.00791 & 0.00606 & -1.30628 &0.19146\\
		$\sigma_{S(4)}$ & 0.15400 & 0.00296 & 51.94285 &0.00000\\
		\multicolumn{5}{l}{Prices of risk of real short rate, inflation, and equity}\\
		$\Lambda_{0(1)}$ & 0.02345 &--- &--- &--- \\
		$\Lambda_{0(2)}$ & -0.17070 & 0.08657 & -1.97194 &0.04862\\
		$\Lambda_{0(4)}$ & 0.28577 &--- &--- &--- \\
		$\Lambda_{1(1,1)}$ & -9.92620 & 5.89176 & -1.68476 &0.09203\\
		$\Lambda_{1(2,2)}$ & -9.98036 & 4.53914 & -2.19873 &0.02790\\
		$\Lambda_{1(4,1)}$ & -14.04888 &--- &--- &--- \\
		$\Lambda_{1(4,2)}$ & -10.36629 &--- &--- &--- \\
		\hline
	\end{tabular}
	\begin{flushleft}
		{\small
			The parameters in the table are annualized. The parameters with standard errors are directly estimated. The parameters without standard errors are obtained by solving three equations: $\delta_R = \delta_r + \delta_{\pi^e} - \sigma^{\top}_{\Pi}\Lambda_0$, $\sigma^{\top}_S\Lambda_0=\mu_0$, $\sigma^{\top}_S\Lambda_1=\mu_1$.
		}
	\end{flushleft}
\end{table}

We consider a pension member who enters the DC plan at age $22$ and invests until age $67$, so the investment horizon is $T=45$. The pension account is invested in a 3-year nominal bond, a 10-year nominal bond, a 10-year inflation-linked bond, the equity index, and cash; that is, $T_1=3$ and $T_2=T_3=10$. As in \cite{koijen2011optimal}, the growth rate in the real contribution process is specified by
\begin{equation*}
	g_t^R=0.1682-0.00646(x+t)+0.00006(x+t)^2,
	\qquad x=22,
\end{equation*}
which corresponds to the high-school income profile estimates in \cite{cocco2005consumption} and \cite{munk2010dynamic}. All monetary quantities are expressed in normalized real units. We set initial real wealth to $W_0^R=1$ and the initial annualized real contribution rate to $C_0^R=6$.

For individual mortality, we use the male data from the 2022 U.S. period life table published by the Social Security Administration (SSA) \citep{ssa2025periodlifetable}. Following \citet{forfar1988graduation}, we assume that the force of mortality follows a generalized Gompertz--Makeham $GM(s_1,s_2)$ model:
\begin{equation*}
	\lambda_a
	=\sum_{i=1}^{s_1}a_i(a-22)^{i-1}
	+\exp\!\left\{\sum_{i=s_1+1}^{s_1+s_2}a_i(a-22)^{i-s_1-1}\right\},
	\qquad 22\leq a\leq67.
\end{equation*}
We estimate 12 candidate models over the grid $s_1\in\{0,1,2,3\}$ and $s_2\in\{2,3,4\}$ by maximum likelihood. The final model is selected using
\begin{equation*}
	\mathrm{BIC}=k\log n-2\ell,
\end{equation*}
where $\ell$ is the maximized log likelihood, $k=s_1+s_2$ is the number of estimated parameters, and $n$ is the sample size. The minimum BIC is attained by the $GM(2,3)$ specification with estimates reported in Table \ref{tab:mortality_estimate_table}.

For the utility specification, we restrict the objective in \eqref{eq:primal_objective} to the CRRA form
\begin{equation*}
	U(x)=\frac{x^{1-\gamma}}{1-\gamma},
\end{equation*}
and report the cases $\gamma=1.5$ and $\gamma=0.8$ to compare different risk preferences.

\begin{table}[!htbp]
	\caption{Estimation results for the force of mortality.}
	\label{tab:mortality_estimate_table}
	\centering
	\begin{tabular}{c r r r}
		\toprule
		Model & $GM(2,3)$ & BIC & $284{,}634.55$ \\
		\midrule
		Parameter & $a_1$ & $a_2$ & $a_3$ \\
		Value & $1.46136171 \times 10^{-3}$ & $9.70090568 \times 10^{-5}$ & $-1.55868347 \times 10^{1}$ \\
		
		Parameter & $a_4$ & $a_5$ & \\
		Value & $4.31929487 \times 10^{-1}$ & $-4.01561314 \times 10^{-3}$ & \\
		\bottomrule
	\end{tabular}
\end{table}
\FloatBarrier

\subsection{Upper bound and lower bound algorithms}
We compare four numerical approaches. Specifically, Benchmarks 1 and 2 and our proposed Reparameterized neural network approach use the dual control framework. Benchmark 3 is a direct neural network approach without a dual transformation. In Benchmark 1, the SAMS approach of \cite{bick2013solving} uses affine-in-time compensated drift terms and has the ODE solution in Proposition~\ref{explicit_solution_1}. In Benchmark 2, the Deep Pricing Kernel approach of \cite{li2024constrained} uses a neural network to optimize the upper bound in Proposition~\ref{semi_explicit_upperbound}, while regression networks recover the lower-bound policy. Our approach uses a reparameterized neural network for the dual upper bound, nested Monte Carlo for the conditional moments needed for the lower bound, and a Hamiltonian quadratic program (QP) to recover a feasible portfolio. 

We also compare an additional direct neural network solution: Benchmark 3, the Direct-Primal Neural Network approach of \cite{li2019data}. We extend their algorithm to incorporate insurance controls as one dimension of the network output.

\par\vspace{\baselineskip}
\noindent\textbf{Dual control methods (via a dual transformation)}\par
\begin{itemize} 
	\item \textbf{Benchmark 1: SAMS approach}\\
	We adopt the affine form of \cite{bick2013solving} as the benchmark. The transformed process $\widetilde v_t$ is assumed to be componentwise affine in time, with nonnegative endpoint values so that $\widetilde v_t\in\widetilde A$ along the whole time grid. The affine coefficients are selected by minimizing the dual upper bound objective. An explicit upper bound is derived in Proposition~\ref{explicit_solution_1}
	\item \textbf{Benchmark 2: Deep Pricing Kernel approach}\\
    Following \cite{li2024constrained}, we parameterize $\widetilde v_t=\mathcal{N\!N}_{\eta}(t,X_t)$ to construct the upper bound neural network. We map the network output into the nonnegative cone $\widetilde A$ and train the network parameters by minimizing the dual upper-bound objective. We use the same deep pricing kernel regression to estimate the lower-bound conditional moments and recover the primal strategies. Because of the high dimensionality, we modify the original truncated projection into a quadratic programming to recover the candidate strategy. Details can be found in Section~\ref{subsec:four_method_comparison}.
	\item \textbf{Proposed method: Reparameterized neural network approach}\\
	Instead of outputting the five constrained components of $\widetilde v_t$ as in Benchmark 2, the network outputs an unconstrained four-dimensional vector $u_t$, which is the reparameterized compensated terms, and a nonnegative scalar slack $s_t$: $(u_t,s_t)=\mathcal{N\!N}_{\eta}(t,X_t)$. Proposition \ref{proposition:minimal_shift_reparameterization} guarantees that these outputs are converted into a feasible $\widetilde v_t$. The network parameters are trained by minimizing the same dual upper bound objective. The lower bound is recovered through Nested Monte Carlo simulation with quadratic programming to guarantee feasibility. 
\end{itemize}
\noindent\textbf{Direct method}\par
\begin{itemize}
    \item \textbf{Benchmark 3: Direct Neural Network approach}\\
    Following the data-driven idea of \cite{li2019data}, we use one neural network to learn the admissible investment and insurance strategies directly, while the constraints over underlying assets are enforced through a softmax activation function. This approach provides a sub-optimal numerical approximation for the optimal value function and strategies.
\end{itemize}
\subsubsection{Details of Benchmark 1: SAMS approach}

When $v_t=v(t)$, i.e., $v_t$ is a function of $t$, we can derive the explicit solution of the upper bound for the primal problem \ref{primal_problem}. 
\begin{proposition}\label{explicit_solution_1} 
	Suppose that $v_t = v(t)$ and $t \in [0, T]$, then the upper bound of the primal problem \ref{primal_problem} is given by
	\begin{equation}
		\widetilde{J}^S(t,W^R_{v,t},X_t;v) = \frac{1}{1-\gamma}[\widetilde{F}^S_1(t,W^R_{v,t},X_t)]^{1-\gamma}[\widetilde{F}^S_2(t,X_t)]^{\gamma},\label{upper_bound_JR}
	\end{equation}
	where the superscript ``S'' is short for the SAMS method and 
	\begin{eqnarray*}
		&&\widetilde{F}^S_1(t,W^R_{v,t},X_t) = W^R_{v,t} + \int_t^T{_{s-t}p_{x+t}}\widetilde{P}^R(X_t,s-t,s)[C^R_s + \delta(\widetilde{v}_s)]ds,\\
        &&\widetilde{F}^S_2(t,X_t) = \int_t^T {_{s-t}p_{x+t}}\lambda_{x+s}\widetilde{F}^S_3(s-t,s) ds + {_{T-t}p_{x+t}}\widetilde{F}^S_3(T-t,T),\\
        &&\widetilde{P}^R(X_t,\tau,s)=\exp[\widetilde{A}_0(\tau,s)+\widetilde{A}_1^{\top}(\tau)X_t],\\
		&&\widetilde{F}^S_3(\tau,s) = \exp{\left\{\widetilde{\Gamma}_0(\tau,s)+\widetilde{\Gamma}^{\top}_1(\tau,s)X_t+\frac{1}{2}X^{\top}_t\widetilde{\Gamma}_2(\tau)X_t\right\}}.
	\end{eqnarray*}
    Functions $\widetilde{A}_0(\tau)\in \mathbb{R}$, $\widetilde{A}_1(\tau)\in \mathbb{R}^2$, $\widetilde{\Gamma}_0(\tau) \in \mathbb{R}$, $\widetilde{\Gamma}_1(\tau) \in \mathbb{R}^2$, and $\widetilde{\Gamma}_2(\tau) \in \mathbb{R}^2 \times \mathbb{R}^2$ are given by the following ODEs
    \begin{align*}
        &\frac{d \widetilde{A}_1(\tau)}{d \tau} + K_X^\top \widetilde{A}_1(\tau)+\Lambda_1^\top \Sigma_X^\top \widetilde{A}_1(\tau) + e_1 = 0, \\
        &\frac{d \widetilde{A}_0(\tau,s)}{d \tau} - \frac{1}{2} \widetilde{A}_1(\tau)^\top S_X \widetilde{A}_1(\tau) + (\delta_r + v_{0}(s-\tau)) + \widetilde{A}_1(\tau)^\top \Sigma_X (\Lambda_0 + v_{-}(s-\tau) - \sigma_{\Pi}) = 0,\\
        &\frac{d \widetilde{\Gamma}_2(\tau)}{d \tau} - \widetilde{\Gamma}_2(\tau) S_X \widetilde{\Gamma}_2(\tau) - B^{\top}_{\gamma} \widetilde{\Gamma}_2(\tau) - \widetilde{\Gamma}_2(\tau) B_{\gamma} - \frac{1-\gamma}{ \gamma^2} \Lambda_1^{\top} \Lambda_1 = 0,\\
        &\frac{d \widetilde{\Gamma}_1(\tau,s)}{d \tau} - \left( \widetilde{\Gamma}_2(\tau) S_X + B^{\top}_{\gamma}\right) \widetilde{\Gamma}_1(\tau,s) - \widetilde{\Gamma}_2(\tau) \widetilde{b}_{\gamma} - \frac{1-\gamma}{\gamma^2} \Lambda_1^{\top} (\Lambda_0 + v_{-}(s-\tau)- \sigma_{\Pi}) - \frac{1-\gamma}{\gamma} e_1 = 0, \\
        &\frac{d \widetilde{\Gamma}_0(\tau,s)}{d \tau} - \frac{1}{2} \widetilde{\Gamma}_1^{\top}(\tau,s) S_X \widetilde{\Gamma}_1(\tau,s) - \widetilde{\Gamma}_1^{\top}(\tau,s)\widetilde{b}_{\gamma} - \frac{1}{2} \text{Tr}\left\{ \Sigma_X^{\top} \widetilde{\Gamma}_2(\tau) \Sigma_X \right\} - \frac{1-\gamma}{\gamma} (\delta_r + v_{0}(s-\tau)) \\
        &\qquad \quad - \frac{1-\gamma}{2 \gamma^2} [\Lambda_0^{\top} + v_{-}(s-\tau) - \sigma_{\Pi}^{\top}] [\Lambda_0 + v_{-}(s-\tau) - \sigma_{\Pi}] = 0,\\ &\widetilde{A}_1(0)=\widetilde{A}_0(0,s)=\widetilde{\Gamma}_2(0)=\widetilde{\Gamma}_1(0,s)=\widetilde{\Gamma}_0(0,s)=0, ~s\in[t,T],\\
        &B_{\gamma}=-K_X+\frac{1-\gamma}{\gamma}\Sigma_X\Lambda_1,\qquad
	\widetilde{b}_{\gamma}=\frac{1-\gamma}{\gamma}\Sigma_X(\Lambda_0 + v_{-}(t) - \sigma_{\Pi}),\qquad
	S_X=\Sigma_X\Sigma_X^{\top}.
    \end{align*}
    where $\text{Tr}$ is the trace of a matrix. 
    The feasibility of the portfolio strategy is enforced through the following quadratic program (QP), which is established in Appendix~\ref{appendix_8}: \footnote{Euclidean/$L^2$ projection $||y^S_t-\theta||^2$ also gives a feasible policy. However, the quadratic program maintains better optimality; see Section~\ref{subsec:four_method_comparison}.}
    \begin{align}
\widehat\theta_{v,t}^{S,R}
&=
\arg\min_{\theta}
[\theta-(y_{v^*,t}^S)^*]^\top
\Sigma\Sigma^\top
[\theta-(y_{v^*,t}^S)^*], \notag\\[-0.25em]
&\text{subject to}\qquad
\iota_4^\top\theta\leq W^R_t,
\qquad
0\leq\theta\leq W^R_t\iota_4.
\label{eq:method1_hamiltonian_qp}
\end{align}
Here, $(y_{v^*,t}^S)^*$ denotes the candidate trading strategy induced from the minimizer $v^*$ of the upper bound, which follows
\begin{equation}
(y_{v^*,t}^S)^*=(\Sigma^\top)^{-1}\left\{W_{v^*,t}^R\sigma_\Pi
-\frac{\kappa_{v^*,t}}{\gamma}\widetilde{F}^S_1
+\Sigma_X^\top \left[\frac{\widetilde{F}^S_1}{\widetilde{F}^S_2}\frac{\partial \widetilde{F}^S_2}{\partial X^{\top}}-\frac{\partial \widetilde{F}^S_1}{\partial X^{\top}}\right]\right\},
\label{eq:qp_unconstrained_candidate}
\end{equation}
where $\frac{\partial}{\partial X}(\cdot) = (\frac{\partial }{\partial x_1}(\cdot), \frac{\partial}{\partial x_2} (\cdot) )$ is the gradient operator with the column vector $X = (x_1, x_2)^{\top}$. $\frac{\partial}{\partial X^{\top}}(\cdot) = (\frac{\partial }{\partial x_1}(\cdot), \frac{\partial}{\partial x_2} (\cdot) )^{\top}$ is the gradient operator with the row vector $X^{\top} = (x_1, x_2)$. The program selects the feasible portfolio that is close to the Hamiltonian of the candidate trading strategy $y^S_t$. Lastly, the candidate insurance strategy is given by
\begin{equation}
     (I^{S}_{v^*,t})^* =\lambda_{x+t}\left[\widetilde{F}^{S}_1(\widetilde{F}^{S}_2)^{-1}-W^R_{v^*,t}\right]. \label{eq:qp_unconstrained_candidate_insurance}
\end{equation}
\end{proposition}
\begin{remark}
    Our trading strategy $\theta_t=(\theta_{1,t},\theta_{2,t},\theta_{3,t},\theta_{4,t})^{\top}$ is four dimension. Truncating each element of the candidate strategy $y^S_t$ by $[0,W^R_t]$ doesn't guarantee that their sum satisfies $\iota_4^\top\theta\leq W^R_t$. Therefore, we propose the QP above to recover the candidate strategy induced from the upper bound. Moreover, we use the original-market Hamiltonian to recover the candidate strategy since the lower bound is computed in the original market. One can also use the artificial market Hamiltonian, but the relative duality gap will increase to 6.24\% from 2.87\% reported in Table \ref{tab:bounds_methods}. 
\end{remark}

\subsubsection{Details of Benchmark 2 and our proposed method}
The Deep Pricing Kernel and Reparameterized neural network approaches share the semi-explicit dual upper bound formula below, conditional on two moment functions. The former follows the deep-pricing-kernel estimator of \cite{li2024constrained} for policy recovery; the latter evaluates the conditional moments and their state derivatives by nested Monte Carlo, as detailed in Algorithms~\ref{alg:nmc_moments} and~\ref{alg:nmc_lower_bound}.
\begin{proposition}\label{semi_explicit_upperbound}
	Suppose direct compensators $\widetilde v_t=\mathcal{N\!N}_{\eta}(t,X_t)$ or reparameterized compensators $(u_t,s_t)=\mathcal{N\!N}_{\eta}(t,X_t)$, then the upper bound of the primal problem \eqref{eq:primal_problem} is given by
	\begin{equation}
		\widetilde{J}^{NN}(t,W^R_{v,t},X_t;v) = \frac{1}{1-\gamma}\widetilde{F}^{NN}_1(t,W^R_{v,t},X_t)^{1-\gamma}\widetilde{F}^{NN}_2(t,X_t)^{\gamma},\label{J_NN}
	\end{equation}
	where $t \in [0, T]$, the superscript ``$NN$'' is short for ``neural network'', and
	\begin{eqnarray*}
		&&\widetilde{F}^{NN}_1(t,W^R_{v,t},X_t) = W^R_{v,t}+\int_t^T {_{s-t}p_{x+t}}E_t\left[\frac{\pi_{v,s}}{\pi_{v,t}}\right] (C^R_s+\delta(\widetilde v_s)) ds, \\
		&&\widetilde{F}^{NN}_2(t,X_t) = \int_t^{T} {_{s-t}p_{x+t}}\lambda_{x+s}E_t\left[\left(\frac{\pi_{v,s}}{\pi_{v,t}}\right)^{-\frac{1-\gamma}{\gamma}}\right] ds + {_{T-t}p_{x+t}} E_t\left[\left(\frac{\pi_{v,T}}{\pi_{v,t}}\right)^{-\frac{1-\gamma}{\gamma}}\right].
	\end{eqnarray*}
	    Furthermore, the unconstrained artificial-market candidate strategies are
		\begin{eqnarray}
			(y^{NN}_{v^*,t})^* &=& (\Sigma^{\top})^{-1}\left\{W^R_{v^*,t}\sigma_{\Pi}-\frac{\kappa_{v^*,t}}{\gamma}\widetilde{F}^{NN}_1+\Sigma^{\top}_X\left[ \frac{\widetilde{F}^{NN}_1}{\widetilde{F}^{NN}_2}\frac{\partial \widetilde{F}^{NN}_2}{\partial X^{\top}}-\frac{\partial \widetilde{F}^{NN}_1}{\partial X^{\top}} \right]\right\},\label{J_NN_strategy_theta}\\
	        (I^{NN}_{v^*,t})^* &=&\lambda_{x+t}\left[\widetilde{F}^{NN}_1(\widetilde{F}^{NN}_2)^{-1}-W^R_{v^*,t}\right],\label{J_NN_strategy_I}
		\end{eqnarray}
        where $v^*$ is the minimizer of the upper bound.
	\end{proposition}
For the three dual-control approaches, the process used in $\kappa_{v,t}$ and $\pi_{v,t}$ is $v_t$. When a method parameterizes $\widetilde v_t$ directly, the corresponding $v_t$ is recovered from the inverse of \eqref{eq:tilde_v_mapping}. In the constraint set considered here, $\delta(\widetilde v_t)=0$ whenever the generated $\widetilde v_t$ is feasible. The integrals above are evaluated by Monte Carlo simulation and the trapezoidal rule on the same time grid used for the pricing kernel simulation.

In the numerical simulation, we find that the calibrated volatility matrix $\Sigma$ is nearly singular. To address this, we propose the following proposition to parametrize $u_t=\Sigma^{-1}(\widetilde v_{-,t}-\widetilde v_{0,t}\iota_4)$ directly. See its outperformance and explanations in Section \ref{subsec:four_method_comparison}

\begin{proposition}\label{proposition:minimal_shift_reparameterization}
For any progressively measurable process $u_t\in\mathbb{R}^4$ and any nonnegative process $s_t$, define
\begin{equation*}
	m_t=\Big[-\min_{1\leq j\leq 4}(\Sigma u_t)_j\Big]^+,
\end{equation*}
and
\begin{equation*}
	\widetilde v_{0,t}=m_t+s_t,\qquad
	\widetilde v_{-,t}=\Sigma u_t+\widetilde v_{0,t}\iota_4.
\end{equation*}
Then $\widetilde v_t=(\widetilde v_{0,t},\widetilde v_{-,t})\in\widetilde A$ for all $t$. Moreover, $m_t$ is the smallest common shift that makes $\Sigma u_t+\widetilde v_{0,t} \iota_4$ nonnegative with $\widetilde v_{0,t} \geq 0$.
\end{proposition}
\begin{proof}
If $a\geq 0$ and $\Sigma u_t+a\iota_4\geq 0$, then $a\geq-(\Sigma u_t)_j$ for all $j$, hence $a\geq m_t$. This is the intuition for defining $m_t$. By definition, $m_t\geq 0$ and $m_t\geq-(\Sigma u_t)_j$ for every $j$. Hence $\widetilde v_{0,t}\geq 0$ and
\begin{equation*}
	(\widetilde v_{-,t})_j=(\Sigma u_t)_j+m_t+s_t\geq 0,\qquad j=1,\ldots,4.
\end{equation*}
So $\widetilde v_t\in\widetilde A$. 
\end{proof}

For the Reparameterized Neural Network approach, given a trained upper bound optimizer $v^*$, the lower bound problem is solved through a nested Monte Carlo method. The inner loop serves to compute the candidate strategies, and the outer loop computes the lower bound by employing the candidate strategies. Let $t_i=i\Delta t$, $\Delta t =T/N_{\text{steps}}$, and write $Y_i=Y_{t_i}$ for any time-dependent process $Y$. For the inner loop, the algorithm estimates the conditional moment functions
\begin{align*}
	F_{1,i}
	&=W_i^R+ \int_{t_i}^T {_{s-t_i}p_{x+t_i}}E^{v*}_{t_i}\left[\frac{\pi_{v^*,s}}{\pi_{v^*,t_i}}\right]C_s^R\,ds,\\
	F_{2,i}
	&=E_i^{v^*}\left[
	\int_{t_i}^T {}_{s-t_i}p_{x+t_i}\lambda_{x+s}
	\left(\frac{\pi_{v^*,s}}{\pi_{v^*,t_i}}\right)^{p_{\gamma}}ds
	+{}_{T-t_i}p_{x+t_i}
	\left(\frac{\pi_{v^*,T}}{\pi_{v^*,t_i}}\right)^{p_{\gamma}}
	\right],
\end{align*}
where $p_{\gamma}=1-1/\gamma$ and $E_i^{v^*}$ denotes conditional expectation given the outer state under the frozen upper bound optimizer. Algorithm \ref{alg:nmc_moments} describes the inner loop estimating these expectations and their derivatives, and Algorithm \ref{alg:nmc_lower_bound} shows the outer loop simulating the lower bound.

% \noindent\rule{\textwidth}{1pt}
% \vspace{-0.4in}
\begin{algorithmblock}{Inner loop estimator for the conditional moment terms}
\label{alg:nmc_moments}
\vspace{-0.2in}
\noindent\rule{\textwidth}{1pt}
\begin{algorithmic}[1]
\Require Outer state $(W_i^R,X_i)$, trained upper bound process $v^*$, $M_{\mathrm{inner}}$, and $\varepsilon_{\mathrm{FD}}$.
\Ensure $\widehat F_{1,i}$, $\widehat F_{2,i}$ and their derivatives with respect to $X$.
\State Let $p_{\gamma}=1-1/\gamma$. Generate common random numbers for the base state and the four perturbed states $X_i\pm\varepsilon_{\mathrm{FD}}e_1$ and $X_i\pm\varepsilon_{\mathrm{FD}}e_2$.
\State Simulate each inner path under the optimizer $v^*$:
\Statex \[
\begin{aligned}
\pi_{i,j+1}^{(m)}
&=\pi_{ij}^{(m)}
\exp\left\{
-\left(r_j^{(m)}+v_{0,j}^{*,(m)}
+\frac12|\kappa_{v^*,j}^{(m)}|^2\right)\Delta t
+\sqrt{\Delta t}\,(\kappa_{v^*,j}^{(m)})^{\top}\xi_j^{(m)}
\right\},\\
X_{j+1}^{(m)}
&=X_j^{(m)}-K_XX_j^{(m)}\Delta t
+\Sigma_X\sqrt{\Delta t}\,\xi_j^{(m)},\qquad
j=i,\ldots,N_{\text{steps}}-1,
\end{aligned}
\]
\Statex where $(m)$ is the path number in the inner loop, $\pi_{ii}^{(m)}=1$, $\xi_j^{(m)}\sim N(0,I_4)$, and $\kappa_{v^*,j}^{(m)}=-(\Lambda_j^{(m)}+v_{-,j}^{*,(m)}-\sigma_{\Pi})$. Thus $\pi_{ij}^{(m)}$ directly simulates the pricing-kernel ratio $\pi_{v^*,t_j}/\pi_{v^*,t_i}$.
\State With trapezoidal weights $a_{ij}$, compute
\Statex \[
\widehat F_{1,i}
=W_i^R+\frac{1}{M_{\mathrm{inner}}}
\sum_{m=1}^{M_{\mathrm{inner}}}
\sum_{j=i}^{N_{\text{steps}}}
a_{ij}\,{}_{t_j-t_i}p_{x+t_i}\,C_{t_j}^R\,\rho_{ij}^{(m)}\,\Delta t,
\]
\Statex \[
\widehat F_{2,i}
=\frac{1}{M_{\mathrm{inner}}}
\sum_{m=1}^{M_{\mathrm{inner}}}
\left[
\sum_{j=i}^{N_{\text{steps}}}
a_{ij}\,{}_{t_j-t_i}p_{x+t_i}\lambda_{x+t_j}
(\rho_{ij}^{(m)})^{p_{\gamma}}\Delta t
+{_{T-t_i}p_{x+t_i}}(\rho_{i,N_{\text{steps}}}^{(m)})^{p_{\gamma}}
\right],
\]
where $M_{\mathrm{inner}}$ and $N_{\text{steps}}$ are the number of path and time steps for the inner loop, repectively.
\State Approximate $\nabla_X\widehat F_{\ell,i}$, $\ell=1,2$, by central finite differences with common random numbers, for example
\Statex \[
\partial_{X_{1,i}}\widehat F_{\ell,i}
\approx
\frac{\widehat F_{\ell,i}(X_{i}+\varepsilon_{\mathrm{FD}}e_1)
-\widehat F_{\ell,i}(X_{i}-\varepsilon_{\mathrm{FD}}e_1)}
{2\varepsilon_{\mathrm{FD}}},\, \partial_{X_{2,i}}\widehat F_{\ell,i}
\approx
\frac{\widehat F_{\ell,i}(X_{2,i}+\varepsilon_{\mathrm{FD}}e_2)
-\widehat F_{\ell,i}(X_{2,i}-\varepsilon_{\mathrm{FD}}e_2)}
{2\varepsilon_{\mathrm{FD}}},
\]
where $e_1 = [1,0]^{\top}$ and $e_2 = [0,1]^{\top}$.
\State Return $\widehat F_{1,i}$, $\widehat F_{2,i}$, and $\nabla_X\widehat F_{\ell,i}=[\partial_{X_{1,i}}\widehat F_{\ell,i}, \partial_{X_{2,i}}\widehat F_{\ell,i}]^{\top}$, $\ell=1,2$.
\end{algorithmic}
\end{algorithmblock}
\noindent\rule{\textwidth}{1pt}

\begin{algorithmblock}{Outer loop simulation for the lower bound}
\label{alg:nmc_lower_bound}
\vspace{-0.2in}
\noindent\rule{\textwidth}{1pt}
\begin{algorithmic}[1]
\Require Trained upper bound process $v^*$;  outer market paths $\{(\varepsilon_i^{(n)},X_i^{(n)},\Lambda_i^{(n)},r_i^{(n)})\}$, with $v^*(t_i,X_i^{(n)})$ evaluated only for pricing-kernel moment recovery; $N_{\mathrm{LB}}$, $M_{\mathrm{inner}}$, and $\varepsilon_{\mathrm{FD}}$.
\Ensure Lower bound estimate $\widehat{\mathrm{LB}}$ and admissible controls $\widehat\theta^R,\widehat I^R$.
\State Set $W_0^{R,(n)}=w^R_0$, $n=1,\ldots,N_{\mathrm{LB}}$.
\For{$i=0,\ldots,N_{\text{steps}}-1$}
	\For{$n=1,\ldots,N_{\mathrm{LB}}$}
		\State Apply Algorithm \ref{alg:nmc_moments} at $(W_i^{R,(n)},X_i^{(n)})$.
		\State Compute the death benefit and insurance premium:
		\Statex \[
		\widehat M_i^{R,(n)}
		=\frac{\widehat F_{1,i}^{(n)}}{\widehat F_{2,i}^{(n)}},
		\qquad
		\widehat I_i^{R,(n)}
        =
        \lambda_{x+t_i}\left(\widehat M_i^{R,(n)}-W_i^{R,(n)}\right).
		\]
		\State Define
		\Statex \[
		\begin{aligned}
		D_{X,i}^{(n)}
		=\frac{\nabla_X\widehat F_{2,i}^{(n)}}{\widehat F_{2,i}^{(n)}}
		-\frac{\nabla_X\widehat F_{1,i}^{(n)}}{\widehat F_{1,i}^{(n)}}, \qquad y_i^{(n)}
		=\left[W_i^{R,(n)}\sigma_{\Pi}^{\top}
		-\frac{\widehat F_{1,i}^{(n)}}{\gamma}(\kappa_{v^*,i}^{(n)})^{\top}
		+\widehat F_{1,i}^{(n)}(D_{X,i}^{(n)})^{\top}\Sigma_X\right]\Sigma^{-1},
		\end{aligned}
		\]
        where $\kappa_{v^*,i}^{(n)}=-(\Lambda_i^{(n)}+(v^{(n)}_i)^*-\sigma_\Pi)$
		\State  Recover the trading strategy by applying the Hamiltonian-loss quadratic program derived in the proof of Proposition
		\ref{explicit_solution_1} (Appendix \ref{appendix_8}):
		\Statex \[
		\widehat\theta_i^{R,(n)}
		=\arg\min_{\substack{\theta\in \mathcal{K}(W_i^{R,(n)})\\ \theta\geq0}}
		(\theta-y_i^{(n)})^\top
		\Sigma\Sigma^\top
		(\theta-y_i^{(n)}).
		\]
		\Statex Here
		\Statex \[
		\mathcal{K}(W_i^{R,(n)})
		=\left\{\theta\in\mathbb R^4:
		\iota_4^\top\theta\leq W_i^{R,(n)}\right\}.
		\]
		\State Set the remaining wealth in the cash account:
		\Statex \[
		\widehat\alpha_i^{R,(n)}
		=W_i^{R,(n)}-\sum_{k=1}^4\widehat\theta_{k,i}^{R,(n)}\geq0.
		\]
		\State Update the outer real wealth path:
		\Statex \[
		\begin{aligned}
		W_{i+1}^{R,(n)}
		=\bigg[ W_i^{R,(n)}
		&+\left\{
		r_i^{(n)}W_i^{R,(n)}
		+
		\left((\widehat\theta_i^{R,(n)})^{\top}\Sigma
		-W_i^{R,(n)}\sigma_{\Pi}^{\top}\right)
		(\Lambda_i^{(n)}-\sigma_{\Pi})
		+C_i^R-\widehat I_i^{R,(n)}
		\right\}\Delta\\
		&+\left((\widehat\theta_i^{R,(n)})^{\top}\Sigma
		-W_i^{R,(n)}\sigma_{\Pi}^{\top}\right)
		\sqrt{\Delta}\,\varepsilon_i^{(n)}
		\bigg]^+.
		\end{aligned}
		\]
	\EndFor
\EndFor
\State Return
\Statex \[
\widehat{\mathrm{LB}}
=\frac{1}{N_{\mathrm{LB}}}\sum_{n=1}^{N_{\mathrm{LB}}}
\left[
\sum_{i=0}^{N_{\text{steps}}}
a_i\,{}_{t_i}p_x\lambda_{x+t_i}U(\widehat M_i^{R,(n)})\Delta
+{}_{T}p_xU(W_{N_{\text{steps}}}^{R,(n)})
\right],
\]
where $a_i$ are the trapezoidal weights on the outer time grid.
\end{algorithmic}
\end{algorithmblock}
\noindent\rule{\textwidth}{1pt}
\FloatBarrier
The positive part operator in the wealth update is only a numerical safeguard against time discretization error. In all 10,000 paths in the lower bound simulation, the wealth updates remained nonnegative before applying this operator.

\subsection{Comparison of the four approaches}\label{subsec:four_method_comparison}

All four approaches use 10,000 paths and 210 time steps for training and evaluation. All trainable models are optimized using Adam for 2,000 iterations at a constant learning rate of $5\times10^{-4}$. All neural networks have three hidden layers of 32 neurons with ReLU activations. B1 optimizes affine parameters directly without any neural network. B2 additionally trains two conditional moment regressors for lower bound estimation. Our proposed method uses 40 inner paths per outer state for nested Monte Carlo simulation. Except for B3, which enforces portfolio constraints through a softmax output, the other three methods impose lower bound portfolio constraints through Hamiltonian quadratic programs solved by OSQP package, with absolute and relative tolerances of $10^{-9}$ and a maximum of 10,000 iterations.

Table \ref{tab:bounds_methods} compares the four approaches. B3 (Direct Neural Network) does not use the dual control framework and therefore reports only one value estimate. Its gap and welfare loss entries are left blank intentionally. 

\begin{table}[htbp]
\centering
\caption{Performance comparison across four approaches}
\label{tab:bounds_methods}
\begin{threeparttable}
\resizebox{\textwidth}{!}{%
\begin{tabular}{lcccc}
\toprule
\multicolumn{5}{c}{$\gamma=1.5$}\\
\midrule
 & B1 (affine) & B2 (NN+DPK) & B3 (NN) & M\\
Parametrization & $\widetilde v_t=a+bt$ & $\widetilde v_t=\mathcal{N\!N}(t,X_t)$ & $\theta_t=\mathcal{N\!N}(t,W_t,X_t)$ & $(u_t,s_t)=\mathcal{N\!N}(t,X_t)$\\
\midrule
Upper bound & $-0.015807$ & $-0.027245$ & \multirow{2}{*}{$-0.052975$} & $-0.050946$\\
Lower bound & $-0.054741$ & $-0.052847$ & & $-0.052448$\\
Absolute gap & $0.038934$ & $0.025601$ & & $0.001503$\\
Relative gap (\%) & $71.12$ & $48.44$ & & $2.87$\\
Welfare loss & $0.916615$ & $0.734201$ & & $0.056485$\\
Premium shape & Hump & Approx. hump & Monotonic & Hump\\
Time elapsed (secs) & $371.07$ & $868.16$ & $432.05$ & $870.62$\\
\midrule
\multicolumn{5}{c}{$\gamma=0.8$}\\
\midrule
 & B1 (affine) & B2 (NN+DPK) & B3 (NN) & M\\
Parametrization & $\widetilde v_t=a+bt$ & $\widetilde v_t=\mathcal{N\!N}(t,X_t)$ & $\theta_t=\mathcal{N\!N}(t,X_t)$ & $(u_t,s_t)=\mathcal{N\!N}(t,X_t)$\\
\midrule
Upper bound & $66.634743$ & $48.007214$ & \multirow{2}{*}{$22.393429$} & $22.707026$\\
Lower bound & $22.458890$ & $22.463536$ & & $22.491808$\\
Absolute gap & $44.175852$ & $25.543678$ & & $0.215217$\\
Relative gap (\%) & $66.30$ & $53.21$ & & $0.95$\\
Welfare loss & $0.995651$ & $0.977569$ & & $0.046500$\\
Premium shape & Hump & Hump & Approx. hump & Hump\\
Time elapsed (secs) & $378.31$ & $919.25$ & $460.33$ & $874.75$\\
\bottomrule
\end{tabular}}
\begin{flushleft}\footnotesize
Notes: B1, B2, B3, and M denote the SAMS, Deep Pricing Kernel, Direct Neural Network, and Reparameterized Network-Nested Monte Carlo approaches, respectively. The absolute gap is $\mathrm{UB}-\mathrm{LB}$; the relative gap is $100(\mathrm{UB}-\mathrm{LB})/\max\{|\mathrm{UB}|,|\mathrm{LB}|\}$. Welfare loss is $1-(\mathrm{LB}/\mathrm{UB})^{1/(1-\gamma)}$. Time elapsed is the total H800 GPU runtime.
\end{flushleft}
\end{threeparttable}
\end{table}

The Reparameterized Neural Network approach gives the tightest dual bounds for both risk preferences. Its relative gap is $2.87\%$ for $\gamma=1.5$, compared with $71.12\%$ for the SAMS approach and $48.44\%$ for the Deep Pricing Kernel approach; for $\gamma=0.8$, the corresponding relative gaps are $0.95\%$, $66.30\%$, and $53.21\%$.  The reported Direct Neural Network utilities, $-0.052975$ and $22.393429$, are below the proposed-method lower estimates, $-0.052448$ and $22.491808$, respectively, and therefore lie outside the reported intervals. Our lower bound exceeds the Direct Neural Network utilities, showing that our candidate strategy outperforms the Direct Neural Network approach. This is because our candidate strategy, induced from the upper bound, retains more structural information. Directly using a neural network to study complex forms such as \eqref{J_NN_strategy_theta} and \eqref{J_NN_strategy_I} is inaccurate.

\begin{figure}[htbp]
    \centering
    \subfigure[$\gamma=1.5$]{\includegraphics[width=0.47\textwidth]{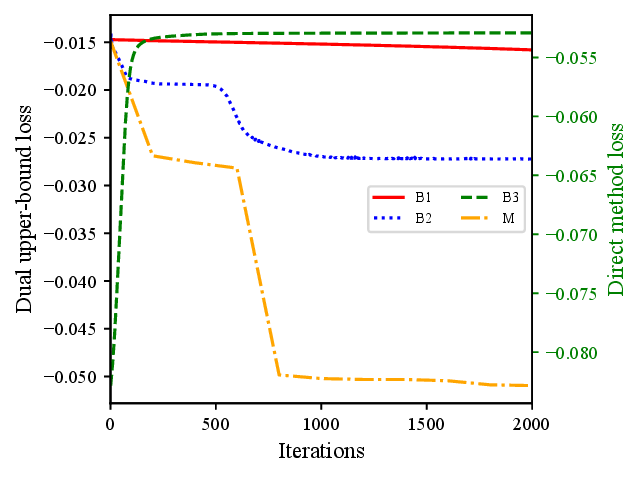}}
    \subfigure[$\gamma=0.8$]{\includegraphics[width=0.47\textwidth]{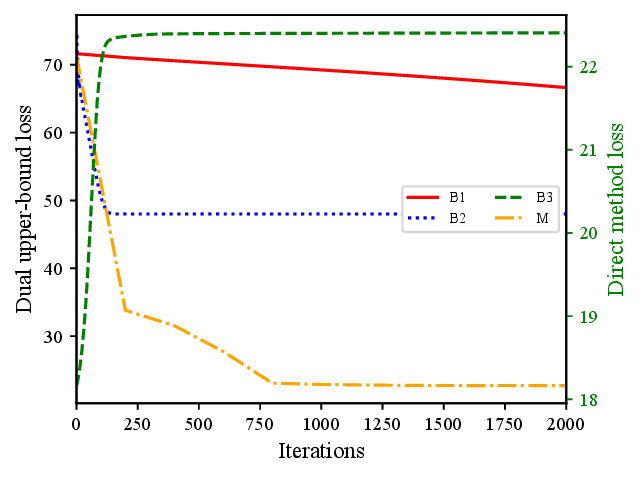}}
    \caption{Loss curves for different methods}
    \label{fig:ub_loss}
\end{figure}

Figure~\ref{fig:ub_loss} shows the loss curves for four methods. The training objectives of all four methods stabilize within the $2{,} 000$ iterations, suggesting that the values reported in Table~\ref{tab:bounds_methods} are close to the final optimal values.

For model calibration, the condition number of the market volatility matrix is $\operatorname{cond}_2(\Sigma)=29.54$, indicating that $\Sigma$ is near singular. Furthermore, by inverse transform \eqref{v_inverse_transform}, the upper bound network's output $\widetilde v_t$ is first differenced and then multiplied by $\Sigma^{-1}$ before entering the pricing kernel. When inverse mapping is ill-conditioned, errors in $\widetilde v_t$ can generate large changes in the compensated drift terms and destabilize gradient-based optimization. This motivates us to parametrize $u_t=\Sigma^{-1}(\widetilde v_{-,t}-\widetilde v_{0,t}\iota_4)$ directly and use the minimal shift construction in Proposition~\ref{proposition:minimal_shift_reparameterization} only to restore feasibility. This change improves the stability of upper bound optimization. The other improvement comes from the Hamiltonian quadratic programming used for lower bound strategy recovery. \cite{li2024constrained} used a truncation to enforce the constraints. However, when the asset dimension expands to four, truncation alone cannot guarantee that the strategy sum is less than or equal to the wealth level.  In a separate comparison for $\gamma=1.5$, holding the other components fixed, the Hamiltonian quadratic program increased the reported lower estimate from $-0.067666$ under Euclidean/$L^2$ projection to $-0.052448$, reducing the relative gap from $24.71\%$ to $2.87\%$.  Furthermore, the nested Monte Carlo approach and the deep pricing kernel regression approach yield comparable results for lower bound conditional moment estimation under otherwise identical conditions. For $\gamma=1.5$, their relative gaps are $2.865316\%$ and $2.810315\%$, respectively; for $\gamma=0.8$, the corresponding gaps are $0.947800\%$ and $0.926117\%$. Nested Monte Carlo is based entirely on simulation and requires no neural network design or hyperparameter tuning. It is therefore be easier to adapt to other models.

\begin{figure}[htbp]
    \centering
    \subfigure[$\gamma=1.5$]{\includegraphics[width=0.47\textwidth]{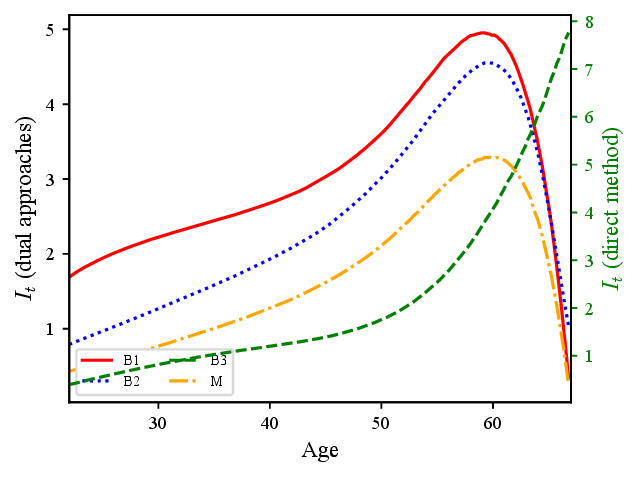}}
    \subfigure[$\gamma=0.8$]{\includegraphics[width=0.47\textwidth]{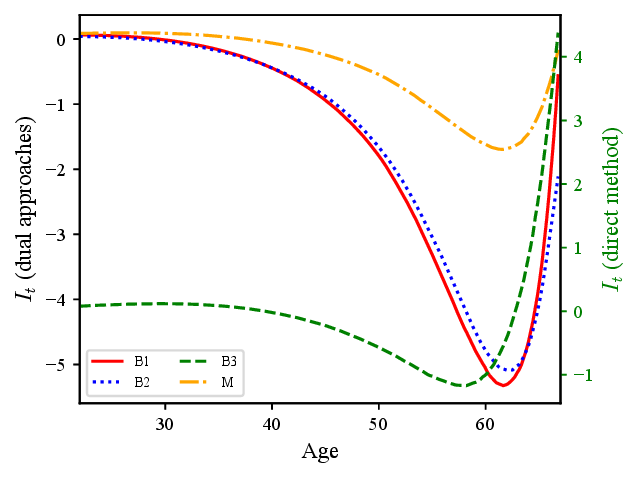}}
    \caption{Expected insurance premium by approach.}
    \label{fig:insurance_profiles_methods}
\end{figure}

Finally, the direct neural network approach fails to recover the hump shape in life insurance purchases, as shown for $\gamma=1.5$ in Figure~\ref{fig:insurance_profiles_methods}. When insurance decisions and asset allocations are optimized directly within the same neural network, the much smaller value of the insurance premium can prevent the network from learning an appropriate insurance policy. The learned premium increases monotonically, although the premium should decline to zero as retirement approaches and the remaining contribution horizon vanishes. In contrast, a dual-control method derives the semi-explicit solutions for the value function and optimal strategies under the general case. More structural information is saved for both the trading and insurance strategies. Neural networks will struggle to learn such a complex strategy form, such as \eqref{J_NN_strategy_theta} and \eqref{J_NN_strategy_I}.

\FloatBarrier

\subsection{Results for the Reparameterized Neural Network approach}\label{subsec:numerical_method3}
\begin{figure}[htbp]
	\centering
	\subfigure[$\gamma=1.5$]{\includegraphics[width=0.47\textwidth]{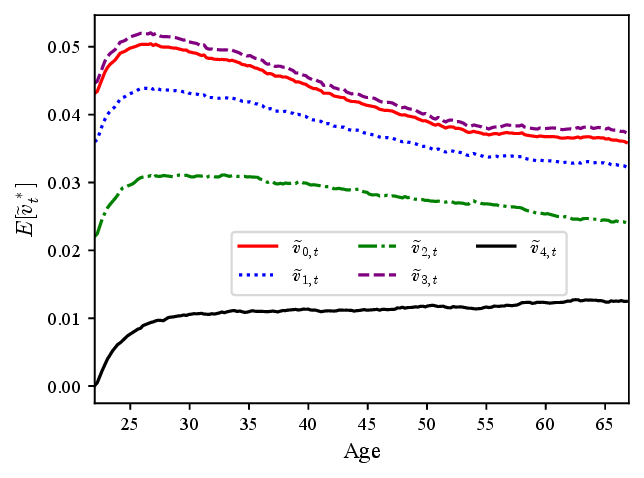}}
	\subfigure[$\gamma=0.8$]{\includegraphics[width=0.47\textwidth]{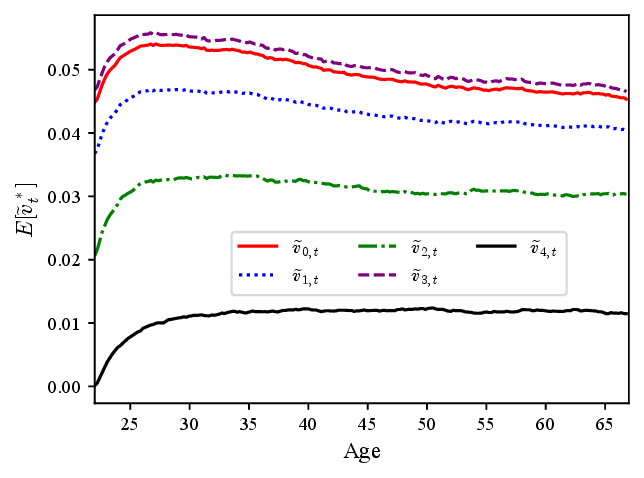}}\\
    [0.8em]
	\subfigure[$\gamma=1.5$]{\includegraphics[width=0.47\textwidth]{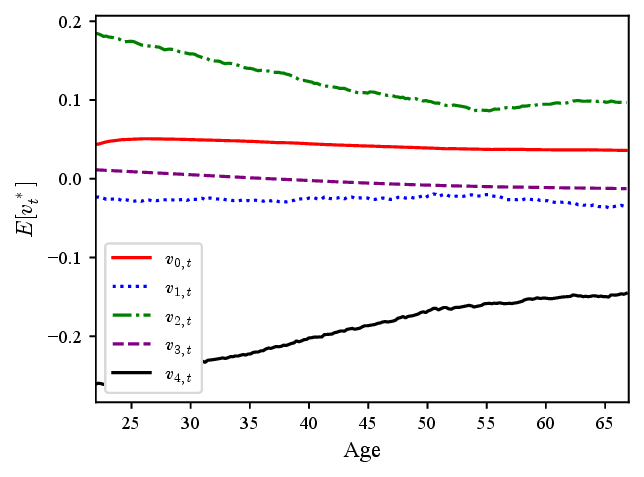}}
	\subfigure[$\gamma=0.8$]{\includegraphics[width=0.47\textwidth]{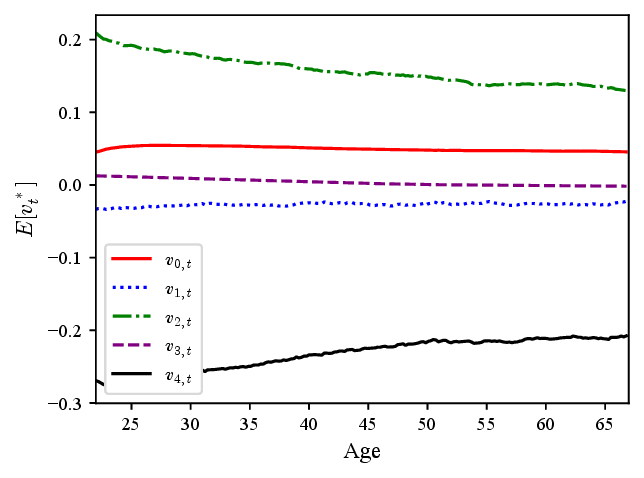}}
	\caption{Expected compensated drift terms.}
	\label{fig:v_components}
\end{figure}

This section presents and explains the
numerical results for the proposed Reparameterized Neural Network approach. Figure~\ref{fig:v_components} reports the expected compensated drift terms that determine the pricing kernel. The component $v_0$ shifts the real short rate, while $v_1$ to $v_4$ adjust the drift terms of the four risky assets. Positive values in $v_0$ and $v_2$ and the significant negative value in $v_4$ indicate that the artificial market adjusts investment opportunities in directions that discourage constraint violations. $v_1$ and $v_3$ are both close to zero, meaning that only small adjustments are needed in those directions. These drift adjustments are consistent with the portfolio constraints imposed in the calibrated model.

\begin{figure}[htbp]
	\centering
	\subfigure[$\gamma=1.5$]{
    \includegraphics[width=0.47\textwidth]{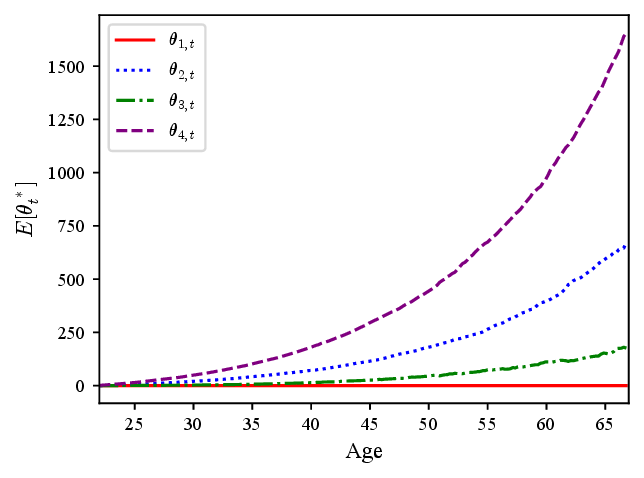}
    \label{fig:theta_and_wealth_1}
    }
	\subfigure[$\gamma=0.8$]{
    \includegraphics[width=0.47\textwidth]{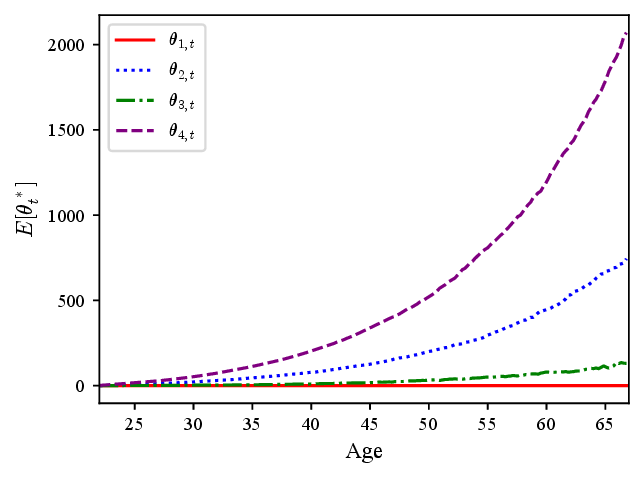}
    \label{fig:theta_and_wealth_2}
    }
	\subfigure[$\gamma=1.5$]{
    \includegraphics[width=0.47\textwidth]{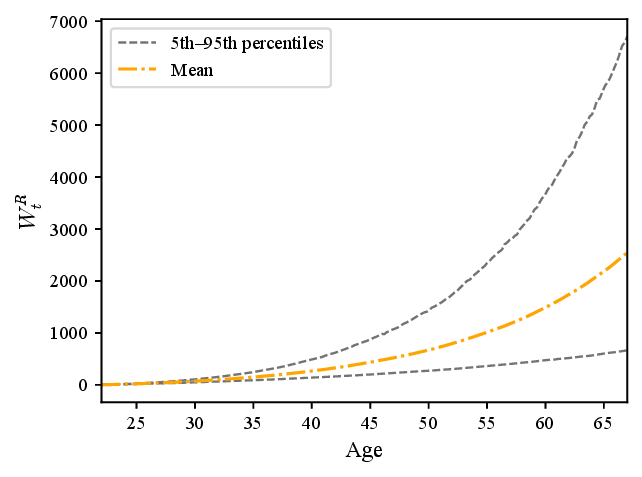}
    \label{fig:theta_and_wealth_3}
    }
	\subfigure[$\gamma=0.8$]{
    \includegraphics[width=0.47\textwidth]{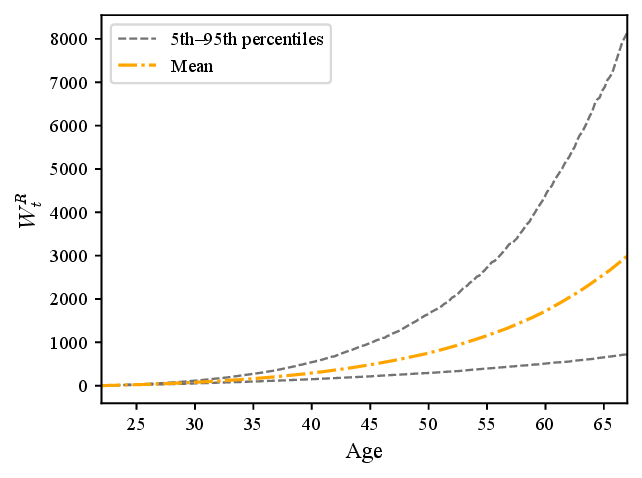}
    \label{fig:theta_and_wealth_4}
    }
	\caption{Optimal wealth process and expected trading strategies.}
	% \label{fig:theta_components}
    \label{fig:theta_and_wealth}
\end{figure}

Figures \ref{fig:theta_and_wealth_1} and \ref{fig:theta_and_wealth_2} shows the expected portfolio holdings implied by the Hamiltonian-QP lower bound policy, with $\theta_1$ to $\theta_4$ representing the real amount invested in the short-term nominal bond, long-term nominal bond, inflation-linked bond, and stock index. As the individual ages and wealth accumulates, he or she has more purchasing power to increase the positions, especially in the stock index. Near retirement, the expected stock position is approximately $1.65\times10^3$ under $\gamma=1.5$ and $2.06\times10^3$ under $\gamma=0.8$.  The expected short-term nominal-bond holding remains close to zero near retirement under both risk-aversion parameters. With no short-selling constraint, the pension member cannot short the short-term nominal bond and purchase other products, which shows the validity of our method. Figures~\ref{fig:theta_and_wealth_3} and \ref{fig:theta_and_wealth_4} summarize the wealth process of our method. Mean wealth rises over the accumulation period, while the $90\%$ confidence interval reflects increasing dispersion across simulated paths. In the normalized units defined in Section~\ref{subsec:model_calibration}, expected terminal wealth is approximately $2.54\times10^3$ and $2.98\times10^3$ under $\gamma=1.5$ and $\gamma=0.8$, respectively.

\begin{figure}[p]
	\centering
	\subfigure[Insurance premium, $\gamma=1.5$]{\includegraphics[width=0.47\textwidth]{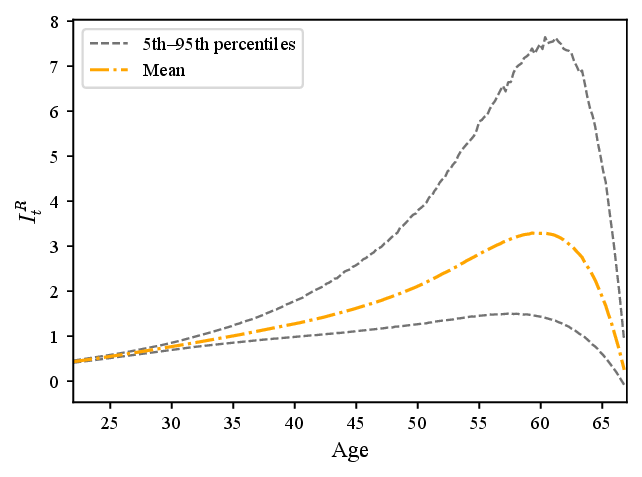}}
	\subfigure[Insurance premium, $\gamma=0.8$]{\includegraphics[width=0.47\textwidth]{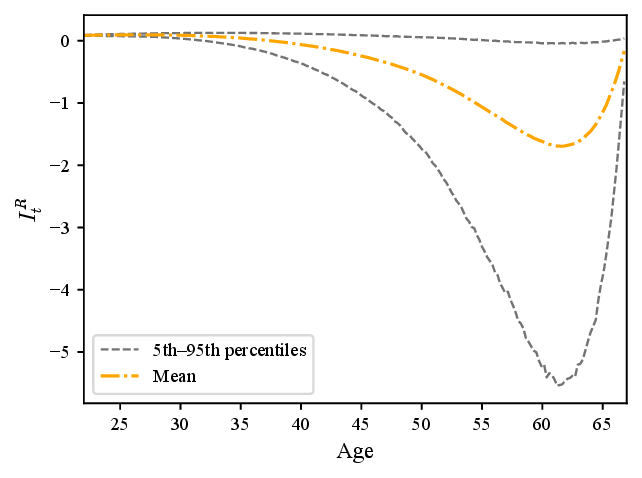}}\\[0.8em]
	\subfigure[Insurance payout, $\gamma=1.5$]{\includegraphics[width=0.47\textwidth]{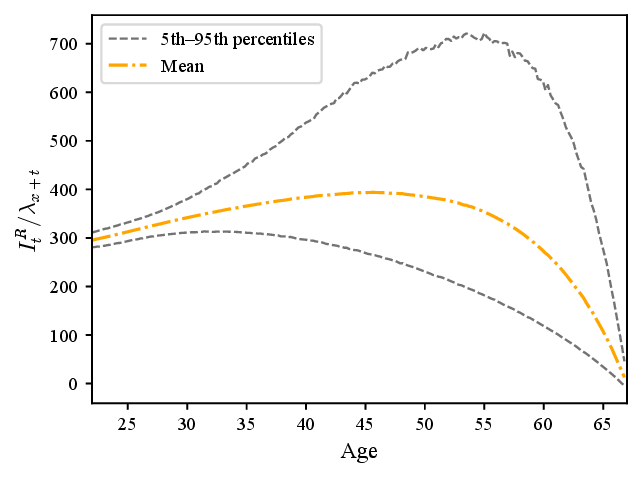}}
	\subfigure[Insurance payout, $\gamma=0.8$]{\includegraphics[width=0.47\textwidth]{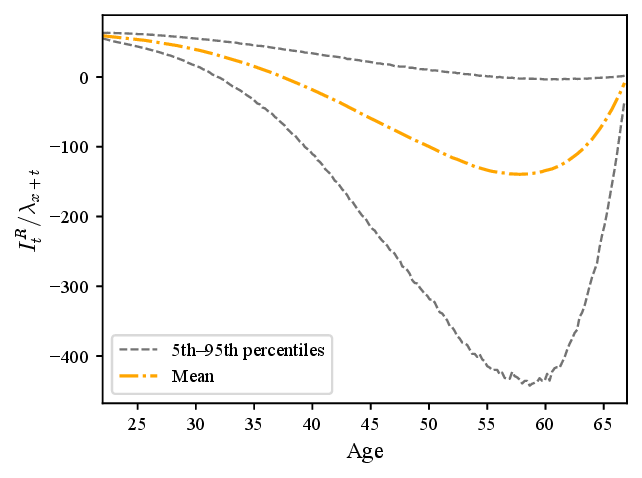}}
	\caption{Expected insurance premium and payout.}
	\label{fig:insurance_premium_payout}
\end{figure}

Differences in risk preference also influence insurance positions. As shown in Figure~\ref{fig:insurance_premium_payout}, when the individual is more risk-averse, the life-insurance premium follows a hump shape and peaks shortly before retirement. This is consistent with the life insurance literature: insurance protects future contributions and accumulated wealth at early ages; as the individual approaches retirement, the remaining contribution horizon shrinks and demand declines. When the individual is less risk-averse with $\gamma=0.8$, the optimal insurance strategy becomes negative after the early accumulation stage. $I^R_t<0$ is usually interpreted as the annuity demand due to the opposite cash flow between the life insurance and annuity (see \cite{hambel2022solving} and \cite{wang2024optimal}). This means the less risk-averse pension member doesn't need a death benefit but an alive payment to invest in risky assets. Moreover, this annuity demand also follows a hump shape and peaks near retirement.

% \begin{figure}[htbp]
% 	\centering
% 	\subfigure[$\gamma=1.5$]{\includegraphics[width=0.47\textwidth]{Figures/gamma1.5/wealth.eps}}
% 	\subfigure[$\gamma=0.8$]{\includegraphics[width=0.47\textwidth]{Figures/gamma0.8/wealth.eps}}
% 	\caption{Optimal wealth process.}
% 	\label{fig:wealth}
% \end{figure}

\subsection{Comparison with unconstrained solution}
We further compare the Reparameterized Neural Network approach with the unconstrained solution derived in Proposition 3.1 in \cite{li2024optimal}. We simulate the unconstrained solution with $N_{\text{paths}}=10000$ Monte Carlo paths and $N_{\text{steps}}=210$ time steps. 

\begin{figure}[p]
	\centering
	\subfigure[Wealth]{\includegraphics[width=0.47\textwidth]{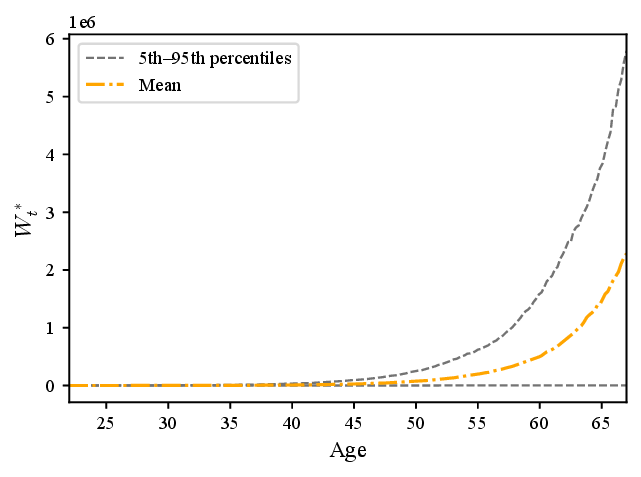}\label{fig:unconstrained_wealth}}
	\subfigure[Portfolio holdings]{\includegraphics[width=0.47\textwidth]{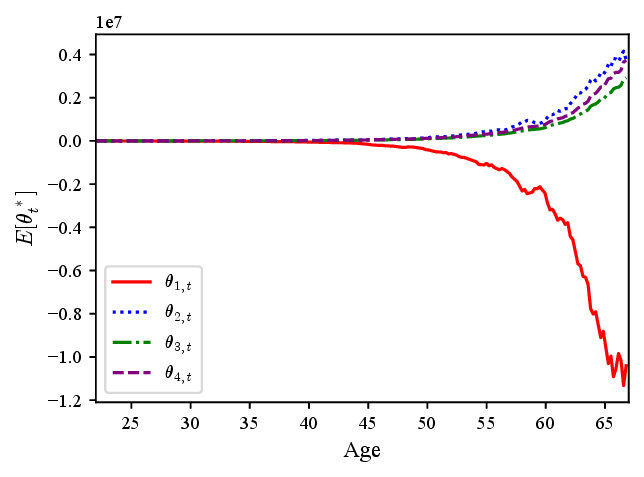}\label{fig:unconstrained_theta}}\\[0.8em]
	\subfigure[Insurance premium]{\includegraphics[width=0.47\textwidth]{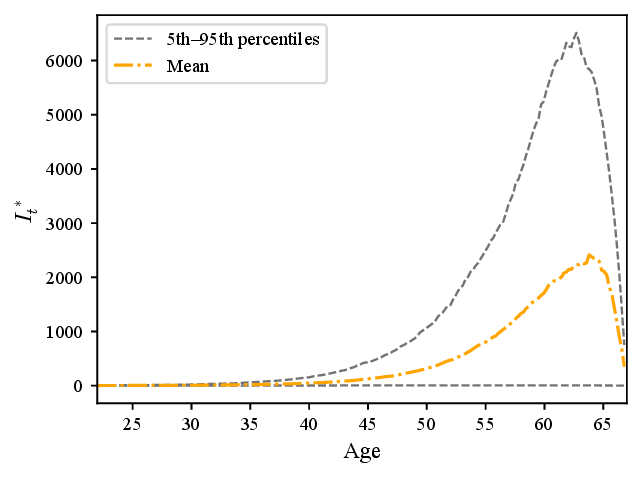}\label{fig:unconstrained_insurance_premium}}
	\subfigure[Insurance payout]{\includegraphics[width=0.47\textwidth]{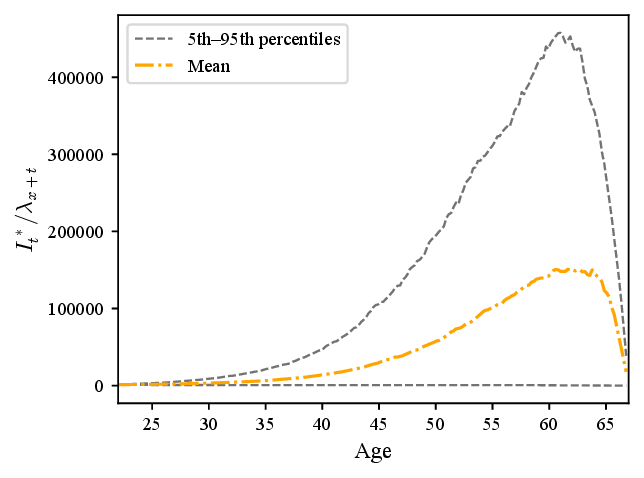}\label{fig:unconstrained_insurance_payout}}
	\caption{Unconstrained analytical solution under $\gamma=1.5$.}
	\label{fig:unconstrained_solution}
\end{figure}

The unconstrained solution provides a useful analytical benchmark, but Figure \ref{fig:unconstrained_solution} also shows its limitation. Without portfolio constraints, the optimal strategy uses extremely aggressive leveraged long and short positions. The participant shorts the short-term nominal bond substantially to finance large long positions in other financial instruments and insurance, as shown in Figures \ref{fig:unconstrained_theta} and \ref{fig:unconstrained_insurance_premium}. These positions violate the portfolio-mixed constraints imposed in our model and pension regulation that ``lending of money or other extension of credit is prohibited in a 401(k) pension fund'' (see \cite{IRC4975}). 

Moreover, the unconstrained solution has limited practical meaning. Compared with the results in Section \ref{subsec:numerical_method3}, the unconstrained benchmark is of much larger magnitude. Expected terminal wealth under $\gamma=1.5$ is approximately $2.54\times 10^3$ for the constrained policy and $2.28\times 10^6$ for the unconstrained benchmark. Without a large short bond position, the participant cannot finance equally large risky holdings and insurance purchases. Consequently, wealth, asset holdings, and insurance positions is enlarged by several orders of magnitude. The results above show that the constrained solution is not a small perturbation of the unconstrained one. Although the unconstrained solution is mathematically tractable, its practical guidance is limited.

This comparison highlights the importance and necessity of incorporating trading constraints into portfolio management problems, especially for pension funds.

\section{Conclusion}\label{Conclusion}

This paper studies a constrained portfolio optimization problem for defined contribution pension plan management. Our financial market is modeled by a two-factor model with a stochastic real interest rate and expected inflation. We allow the pension participant to invest in nominal bonds, an inflation-linked bond, a stock index, and cash. The individual can also purchase term life insurance to hedge against premature death risk before retirement. We introduce a convex trading constraint set to incorporate real-life trading constraints such as no-short-selling and no-borrowing constraints. To solve the problem, we introduce an artificial market and formulate the problem as a dual control problem. We use a reparameterized neural network to minimize the upper bound, then use nested Monte Carlo to estimate the conditional moments and Hamiltonian quadratic programming to maximize the lower bound. Across the two risk preferences, this approach achieves relative gaps of $2.87\%$ and $0.95\%$, substantially smaller than the affine and deep pricing kernel benchmarks. The constrained strategies avoid the large leveraged positions of the unconstrained solution, suggesting the importance of considering trading constraints in practice.

Several extensions are possible for future research. One may introduce transaction costs and more detailed borrowing limits to study implementation. Moreover, one can also replace the continuously traded insurance position with discrete contracts. Lastly, a further extension is to introduce stochastic mortality and examine how mortality risk interacts with financial-market incompleteness and portfolio constraints. These extensions would draw down research to practice and provide more insightful conclusions for pension management.

% \clearpage

\bibliographystyle{apalike}
\bibliography{ref}

\appendix

\section{Proof of Theorem \ref{theorem:static_budget_constraint}}\label{appendix_1}
\begin{proof}
\textit{Only if direction.}
Suppose that $(M^R,W_T^R)\in G_+^*$ is $A$-feasible. Then there exist
$(\alpha^R,\theta^R,I^R)\in\Theta$ and a non-negative, non-decreasing process
$D^R$ such that the real wealth \eqref{eq:real_wealth} holds.
Fix $v\in\mathcal{N}^*$ and write
\[
	\eta_t^R=(\theta_t^R)^{\top}\Sigma-W_t^R\sigma_{\Pi}^{\top}.
\]
By \eqref{eq:kappa_v} and \eqref{eq:Zv},
\[
	dZ_t=dZ_{v,t}+\kappa_{v,t}\,dt
	=dZ_{v,t}-(\Lambda_t+v_{-,t}-\sigma_\Pi)\,dt.
\]
Substituting this identity into the real wealth dynamics
\eqref{eq:real_wealth} gives
\begin{align*}
	dW_t^R
	={}&\left[r_tW_t^R+\eta_t^R(\Lambda_t-\sigma_\Pi)\right]dt
	-\eta_t^R(\Lambda_t+v_{-,t}-\sigma_\Pi)dt\\
	&+\eta_t^R\,dZ_{v,t}+C_t^Rdt-I_t^Rdt-dD_t^R\\
	={}&\left[r_tW_t^R-\eta_t^Rv_{-,t}+C_t^R-I_t^R\right]dt
	+\eta_t^R\,dZ_{v,t}-dD_t^R.
\end{align*}
By Ito's formula and the real death benefit \eqref{real_death_benefit}, we have
\begin{align*}
	d({_t p_x}\beta_{v,t}W_t^R)
	={}&{_t p_x}\beta_{v,t}\,dW_t^R
	+W_t^R\,d({_t p_x}\beta_{v,t})\\
	={}&{_t p_x}\beta_{v,t}
	\left[-v_{0,t}W_t^R-\eta_t^Rv_{-,t}
	+C_t^R-I_t^R-\lambda_{x+t}W_t^R\right]dt\\
	&+{_t p_x}\beta_{v,t}\eta_t^R\,dZ_{v,t}
	-{_t p_x}\beta_{v,t}\,dD_t^R\\
	={}&{_t p_x}\beta_{v,t}
	\left[-v_{0,t}W_t^R-\eta_t^Rv_{-,t}
	+C_t^R-\lambda_{x+t}M_t^R\right]dt\\
	&+{_t p_x}\beta_{v,t}\eta_t^R\,dZ_{v,t}
	-{_t p_x}\beta_{v,t}\,dD_t^R.
\end{align*}
By $W_t^R=\alpha_t^R+\iota_4^\top\theta_t^R$ and
\eqref{eq:tilde_v_mapping}, we can complete a form of supporting function
\begin{align*}
	-v_{0,t}W_t^R-\eta_t^Rv_{-,t}
	={}&-v_{0,t}W_t^R-(\theta_t^R)^\top\Sigma v_{-,t}
	+W_t^R\sigma_\Pi^\top v_{-,t}\\
	={}&-W_t^R\widetilde v_{0,t}
	-(\theta_t^R)^\top\Sigma v_{-,t}\\
	={}&-\alpha_t^R\widetilde v_{0,t}
	-(\theta_t^R)^\top
	\left(\Sigma v_{-,t}+\widetilde v_{0,t}\iota_4\right)\\
	={}&-\alpha_t^R\widetilde v_{0,t}
	-(\theta_t^R)^\top\widetilde v_{-,t}.
\end{align*}
This transformation is invertible because $\Sigma$ is
nonsingular:
\begin{equation}
	v_{-,t}=\Sigma^{-1}
	(\widetilde v_{-,t}-\widetilde v_{0,t}\iota_4),
	\qquad
	v_{0,t}=\widetilde v_{0,t}
	+\sigma_\Pi^\top\Sigma^{-1}
	(\widetilde v_{-,t}-\widetilde v_{0,t}\iota_4). \label{v_inverse_transform}
\end{equation}
Both maps are fixed linear transformations. Hence, by the triangle inequality,
$v\in L^2$ if and only if $\widetilde v\in L^2$.
Consequently,
\begin{align*}
	d({_t p_x}\beta_{v,t}W_t^R)
	={}&{_t p_x}\beta_{v,t}
	\left[-\alpha_t^R\widetilde v_{0,t}
	-(\theta_t^R)^{\top}\widetilde v_{-,t}
	+C_t^R-\lambda_{x+t}M_t^R\right]dt\\
	&+{_t p_x}\beta_{v,t}\eta_t^R\,dZ_{v,t}
	-{_t p_x}\beta_{v,t}\,dD_t^R .
\end{align*}
Integrating from $0$ to $t$ gives
\begin{align}
	{_t p_x}\beta_{v,t}W_t^R
	+\int_0^t {_s p_x}\beta_{v,s}\lambda_{x+s}M_s^R\,ds =w^R_0 &+\int_0^t{_s p_x}\beta_{v,s}
	\left[-\alpha_s^R\widetilde v_{0,s}
	-(\theta_s^R)^{\top}\widetilde v_{-,s}
	+C_s^R\right]ds \notag\\
	&+\int_0^t {_s p_x}\beta_{v,s}H_s^R\,dZ_{v,s}
	-\int_0^t {_s p_x}\beta_{v,s}\,dD_s^R . \label{eq:proof_static_integrated}
\end{align}
Since $(\alpha_s^R,\theta_s^R)\in A$, the definition of the support function
\eqref{eq:support_function} implies
\[
	-\alpha_s^R\widetilde v_{0,s}
	-(\theta_s^R)^{\top}\widetilde v_{-,s}
	\leq \delta(\widetilde v_s).
\]
Moreover, $\int_0^t {_s p_x}\beta_{v,s}\,dD_s^R\geq0$. Hence \eqref{eq:proof_static_integrated}
yields
\begin{align}
	{_t p_x}\beta_{v,t}W_t^R
	+\int_0^t {_s p_x}\beta_{v,s}\lambda_{x+s}M_s^R\,ds 
	\leq{} w^R_0+\int_0^t {_s p_x}\beta_{v,s}
	\left(C_s^R+\delta(\widetilde v_s)\right)ds
	+\int_0^t {_s p_x}\beta_{v,s}H_s^R\,dZ_{v,s}. \label{eq:proof_static_pre_stop}
\end{align}

Define the stopping times
\[
	\tau_n=T\wedge\inf\left\{t\in[0,T]:
	\int_0^t \left| {_s p_x}\beta_{v,s}H_s^R\right|^2ds\geq n\right\},
	\qquad n\geq1,
\]
with $\inf\emptyset=\infty$. Since $(\alpha^R,\theta^R,I^R)\in\Theta$, we have
$\tau_n\uparrow T$ a.s. The stochastic integral in
\eqref{eq:proof_static_pre_stop} is a $Q_v$-martingale on $[0,\tau_n]$. Setting
$t=\tau_n$ and taking $Q_v$-expectations therefore gives
\begin{align}
	E^{Q_v}\left[
		{_{\tau_n} p _x}\beta_{v,\tau_n}W_{\tau_n}^R
		+\int_0^{\tau_n} {_sp_x}\beta_{v,s}\lambda_{x+s}M_s^R\,ds
	\right]
	\leq
	w^R_0+
	E^{Q_v}\left[
		\int_0^{\tau_n} {_sp_x}\beta_{v,s}
		\left(C_s^R+\delta(\widetilde v_s)\right)ds
	\right]. \label{eq:proof_static_stopped_budget}
\end{align}
Because $M^R\geq0$ and $W_T^R\geq0$, Fatou's lemma and monotone convergence imply
\begin{align*}
	&E^{Q_v}\left[
		{_Tp_x}\beta_{v,T}W_T^R
		+\int_0^T {_sp_x}\beta_{v,s}\lambda_{x+s}M_s^R\,ds
	\right] \leq
	\liminf_{n\to\infty}
	E^{Q_v}\left[
		{_{\tau_n}p_x}\beta_{v,\tau_n}W_{\tau_n}^R
		+\int_0^{\tau_n}{_sp_x}\beta_{v,s}\lambda_{x+s}M_s^R\,ds
	\right].
\end{align*}
For the right-hand side of \eqref{eq:proof_static_stopped_budget}, we don't require this endowment to be finite because of the OU-type stochastic interest rate. This extension to infinity doesn't affect our analysis, since our value function's upper bound is the infimum over $v$. We know at least one measure $Q_0$ has finite endowment from the finite expected future contribution in \eqref{wealth_finite_LB} and the upper boundedness of $\delta$ in Assumption \ref{support_function_assumption}. 

In the present portfolio-mix constraint, $\widetilde v_s\in\widetilde A$ implies $\delta(\widetilde v_s)=0$; equivalently, the displayed formula remains valid with the support-function term included. Letting $n\to\infty$ in
\eqref{eq:proof_static_stopped_budget} gives
\begin{align*}
	E^{Q_v}\left[
		{_T p_x}\beta_{v,T}W_T^R
		+\int_0^T{_t p_x}\beta_{v,t}\lambda_{x+t}M_t^R\,dt
	\right]
	\leq
	w^R_0+
	E^{Q_v}\left[
		\int_0^T {_t p_x}\beta_{v,t}
		\left(C_t^R+\delta(\widetilde v_t)\right)dt
	\right].
\end{align*}
Since $v\in\mathcal{N}^*$ is arbitrary, \eqref{eq:static_budget_constraint} follows.

\textit{If direction.}
Now suppose that \eqref{eq:static_budget_constraint} holds for every
$v\in\mathcal{N}^*$. Let $\mathcal{T}$ be the set of stopping times bounded by $T$.
For $\tau\in\mathcal{T}$, define
\begin{align}
	\widehat W^R_{\tau}
	=\operatorname*{ess\,sup}_{v\in\mathcal{N}^*}
	E_{\tau}^{Q_v}\bigg[
	&\int_{\tau}^{T} {_{s-\tau} p_{x+\tau}}\frac{\beta_{v,s}}{\beta_{v,\tau}}
	\left(\lambda_{x+s}M_s^R-C_s^R-\delta(\widetilde v_s)\right)ds + {_{T-\tau} p_{x+\tau}}\frac{\beta_{v,T}}{\beta_{v,\tau}}
	W_T^R
	\bigg]. \label{eq:proof_static_candidate_wealth}
\end{align}
At $\tau=0$, condition \eqref{eq:static_budget_constraint} gives
\[
	\widehat W^R_0\leq w^R_0.
\]
At $\tau=T$, the definition gives $\widehat W^R_T=W_T^R$. By the non-negativity of $M^R_s$, $W^R_T$, and $-\delta(\widetilde{v})$, we have the following bound from \eqref{eq:proof_static_candidate_wealth} 
\begin{equation*}
    \widehat W^R_{\tau} \geq -E_{\tau}^{Q_v}\bigg[
	\int_{\tau}^{T} {_{s-\tau} p_{x+\tau}}\frac{\beta_{v,s}}{\beta_{v,\tau}}
	C_s^Rds\bigg], ~~\text{for}~~ v\in\mathcal{N}^*.
\end{equation*}
Taking the essential supremum on the right-hand side, we obtain 
\begin{eqnarray*}
    \widehat W^R_{\tau} &\geq& \operatorname*{ess\,sup}_{ v\in\mathcal{N}^*} -E_{\tau}^{Q_v}\bigg[
	\int_{\tau}^{T} {_{s-\tau} p_{x+\tau}}\frac{\beta_{v,s}}{\beta_{v,\tau}}
	C_s^Rds\bigg]=-\operatorname*{ess\,inf}_{ v\in\mathcal{N}^*} E_{\tau}^{Q_v}\bigg[
	\int_{\tau}^{T} {_{s-\tau} p_{x+\tau}}\frac{\beta_{v,s}}{\beta_{v,\tau}}C_s^Rds\bigg]\\
    &>& -E_{\tau}^{Q_0}\bigg[
	\int_{\tau}^{T} {_{s-\tau} p_{x+\tau}}\frac{\beta_{0,s}}{\beta_{0,\tau}}C_s^Rds\bigg] = - \int_{\tau}^T {_{s-\tau} p_{x+\tau}} P^R(X_{\tau}, \tau, s)C^R_s ds,
\end{eqnarray*}
which is the finite lower bound requirement in \eqref{wealth_finite_LB}.

The dynamic programming principle applied to \eqref{eq:proof_static_candidate_wealth}
implies that, for each fixed $v\in\mathcal{N}^*$,
\begin{equation}\label{eq:proof_static_supermartingale}
	H_{v,t}
	=
	{_t p_x}\beta_{v,t}\widehat W^R_t
	+\int_0^t {_s p_x}\beta_{v,s}
	\left(\lambda_{x+s}M_s^R-C_s^R-\delta(\widetilde v_s)\right)ds
\end{equation}
is a $Q_v$-supermartingale. For the $H_{v,t}$ in \eqref{eq:proof_static_supermartingale}, by the Doob-Meyer decomposition and the martingale representation theorem, there exist a non-decreasing process $A_v$ with $A_{v,0}=0$ and a progressively measurable $\mathbb{R}^4$-valued process $\Psi_v$ such that
\[
	H_{v,t}
	=
	\widehat W^R_0+\int_0^t\Psi_{v,s}^{\top}dZ_{v,s}-A_{v,t}.
\]
Since \eqref{eq:proof_static_supermartingale} holds for each fixed $v\in\mathcal{N}^*$, $\widehat W^R_t$ under markets $\mathcal{M}_v$ and $\mathcal{M}_0$ are given by
\begin{align*}
 \widehat W^R_t &=(_tp_x)^{-1}\beta^{-1}_{v,t}\left\{ \widehat W^R_0 + \int_0^t \Psi^{\top}_{v,s} dZ_{v,s} - A_{v,t} - \int_0^t {_sp_x} \beta_{v,s} [\lambda_{x+s}M^R_s - C^R_s-\delta(\widetilde{v}_s)]ds\right\}\\ 
 &=(_tp_x)^{-1}\beta^{-1}_{0,t}\left\{ \widehat W^R_0 + \int_0^t \Psi^{\top}_{0,s} dZ_{0,s} - A_{0,t} - \int_0^t {_sp_x} \beta_{0,s} [\lambda_{x+s}M^R_s - C^R_s-\delta(0)]ds\right\},
\end{align*}
where $\delta(0)=0$. Then by Itô's formula and change of measure \eqref{eq:Zv}, we have 
\begin{align}
    d \widehat W^R_t &= [r_t + v_{0,t} + \lambda_{x+t}]\widehat W^R_t dt + (_tp_x)^{-1}\beta^{-1}_{v,t} \Psi^{\top}_{v,t} [dZ_t + (\Lambda_t + v_{-,t}-\sigma_{\Pi})dt]\notag\\
    &\quad -(_tp_x)^{-1}\beta^{-1}_{v,t} dA_{v,t} - [\lambda_{x+t}M^R_t-C^R_t-\delta(\widetilde{v}_t)]dt \label{wealth_Mv}
\end{align}
under artificial market $\mathcal{M}_v$ and 
\begin{align}
    d \widehat W^R_t &= [r_t + \lambda_{x+t}]\widehat W^R_t dt + (_tp_x)^{-1}\beta^{-1}_{0,t} \Psi^{\top}_{0,t} [dZ_t + (\Lambda_t -\sigma_{\Pi})dt]\notag\\ 
    &\quad -(_tp_x)^{-1}\beta^{-1}_{0,t} dA_{0,t} - [\lambda_{x+t}M^R_t-C^R_t]dt \label{wealth_M0}
\end{align}
under market $\mathcal{M}_0$. Compare terms in \eqref{wealth_Mv} and \eqref{wealth_M0}, we obtain 
\begin{align}
    &\beta^{-1}_{v,t} \Psi^{\top}_{v,t} = \beta^{-1}_{0,t} \Psi^{\top}_{0,t},\notag\\
    &\int_0^t v_{0,s} \widehat{W}^R_s + (_sp_x)^{-1}\beta^{-1}_{v,s} \Psi^{\top}_{v,s}  v_{-,s} - (_sp_x)^{-1}\beta^{-1}_{v,s} dA_{v,s} + \delta(\widetilde{v}_s)ds = -\int_0^t (_sp_x)^{-1}\beta^{-1}_{0,s}dA_{0,s}. \label{compare_drift}
\end{align}
Let 
\begin{align}
    &(_tp_x)^{-1}\beta^{-1}_{0,t} \Psi^{\top}_{0,t} = (_tp_x)^{-1}\beta^{-1}_{v,t} \Psi^{\top}_{v,t} = (\theta^R_t)^{\top}\Sigma - \widehat{W}^R_t \sigma^{\top}_{\Pi},\label{vol_transform}\\ 
    &I^R_t = \lambda_{x+t} (M^R_t - \widehat{W}^R_t), ~~D^R_t = w^R_0-\widehat{W}^R_0+\int_0^t ({_sp_x})^{-1}\beta^{-1}_{0,s} dA_{0,s}.\notag
\end{align}
then the wealth process \eqref{wealth_M0} can be rewritten as
\begin{align*}
    \widehat{W}^R_t &= w^R_0 + \int_0^t \{ r_s \widehat{W}^R_s + [(\theta^R_s)^{\top}\Sigma - \widehat{W}^R_s \sigma^{\top}_{\Pi}](\Lambda_s-\sigma_{\Pi}) \} ds + [(\theta^R_s)^{\top}\Sigma - \widehat{W}^R_s \sigma^{\top}_{\Pi}] dZ_s\\
    &+ \int_0^t [C^R_u - I^R_u]du - D^R_t,
\end{align*}
which is the same as the dynamic budget constraint \eqref{eq:real_wealth}.

It remains to show that $(\alpha_t^R,\theta_t^R)\in A$. By \eqref{vol_transform}, \eqref{eq:tilde_v_mapping}, and $\widehat{W}^R_t=\alpha^R_t+\iota_4^{\top}\theta^R_t$,  \eqref{compare_drift} can be rewritten as
\begin{align}
	\int_0^t
	\left(\alpha_s^R\widetilde v_{0,s}
	+(\theta_s^R)^{\top}\widetilde v_{-,s}
	+\delta(\widetilde v_s)\right)ds
	+\int_0^t\beta_{0,s}^{-1}p_s^{-1}\,dA_{0,s}
	=
	\int_0^t\beta_{v,s}^{-1}p_s^{-1}\,dA_{v,s}
	\geq0. \label{eq:proof_static_constraint_ineq}
\end{align}
The mapping $v\mapsto\widetilde v$ in \eqref{eq:tilde_v_mapping} is one-to-one under
the nonsingularity of $\Sigma$. Moreover, $\widetilde A$ is a cone and $\delta$ is
positively homogeneous. If there were a set of positive product measure on which
\[
	\alpha_t^R u_0+(\theta_t^R)^{\top}u_-+\delta(u)<0
\]
for some $u=(u_0,u_-)\in\widetilde A$, then replacing $u$ by $bu$ with $b>0$ and
letting $b$ be large would contradict \eqref{eq:proof_static_constraint_ineq}. Hence
\[
	\alpha_t^R u_0+(\theta_t^R)^{\top}u_-+\delta(u)\geq0,
	\qquad \forall u\in\widetilde A,
\]
for almost every $(t,\omega)$. By the support-function characterization of a closed convex set, this is equivalent to
$(\alpha_t^R,\theta_t^R)\in A$ a.e. (see Theorem 13.1 in \cite{rockafellar1997convex}). This finishes the proof for the \textit{If direction}, inducing the dynamic budget constraint from the static budget constraint.
\end{proof}

\section{Proof of Proposition \ref{prop:fixed_v_dual}}
\begin{proof}
Fix $v\in\mathcal{N}^*$. By Remark \ref{lemma:dual_utility},
$\widetilde U$ is strictly convex and
$\widetilde U'(\psi\pi_{v,t})=-f(\psi\pi_{v,t})$. Hence, for fixed $v$,
$\psi\mapsto\widetilde J(\psi,v)$ is strictly convex on $(0,\infty)$, and
\begin{align}
	\frac{\partial \widetilde J(\psi,v)}{\partial\psi}
	=E\bigg[
	&-\int_0^T {_t p_x}\lambda_{x+t}\pi_{v,t}
	f(\psi\pi_{v,t})\,dt
	-{_T p_x}\pi_{v,T}f(\psi\pi_{v,T}) +w^R_0+\int_0^T {_t p_x}\pi_{v,t}
	\left(C_t^R+\delta(\widetilde v_t)\right)dt
	\bigg]. \label{eq:proof_fixed_v_derivative}
\end{align}

We now check that this derivative has a unique zero point. Since $f$ is strictly decreasing,
the first two terms in \eqref{eq:proof_fixed_v_derivative} are strictly increasing in
$\psi$. Thus $\partial\widetilde J(\psi,v)/\partial\psi$ is strictly increasing. Moreover,
the Inada condition implies
\[
	f(0+)=\infty,\qquad f(\infty)=0.
\]
Therefore
\[
	\lim_{\psi\downarrow0}
	\frac{\partial\widetilde J(\psi,v)}{\partial\psi}<0,
\]
while
\[
	\lim_{\psi\uparrow\infty}
	\frac{\partial\widetilde J(\psi,v)}{\partial\psi}
	=
	w^R_0+E\left[\int_0^T {_t p_x}\pi_{v,t}
	\left(C_t^R+\delta(\widetilde v_t)\right)dt\right]>0.
\]
By continuity, there exists a unique $\psi_v>0$ such that
\[
	\frac{\partial\widetilde J(\psi_v,v)}{\partial\psi}=0.
\]
Strict convexity then implies that this $\psi_v$ is the unique minimizer of
$\widetilde J(\cdot,v)$.

Substituting $\psi=\psi_v$ into \eqref{eq:proof_fixed_v_derivative} gives the fixed-market budget constraint
\begin{align}
	E\left[
		\int_0^T {_t p_x}\lambda_{x+t}\pi_{v,t}
		f(\psi_v\pi_{v,t})\,dt
		+{_T p_x}\pi_{v,T}f(\psi_v\pi_{v,T})
	\right]
	=
	w^R_0+E\left[\int_0^T {_t p_x}\pi_{v,t}
	\left(C_t^R+\delta(\widetilde v_t)\right)dt\right]. \label{eq:proof_fixed_v_budget}
\end{align}
Let
\[
	M_{v,t}^R=f(\psi_v\pi_{v,t}),
	\qquad
	W_{v,T}^R=f(\psi_v\pi_{v,T}),
\]
Then, we can obtain the associated wealth process $W^R_{v,t}$ in \eqref{eq:wealth_fixed_v} by the fixed-market budget constraint \eqref{eq:proof_fixed_v_budget}, which is a $Q_v$-martingale. Therefore, the free-disposal process $D_t$ is zero under artificial market $\mathcal{M}_v$.
\end{proof}

\section{Proof of Proposition \ref{explicit_solution_1}}\label{appendix_8}

\setcounter{equation}{0}
\renewcommand{\theequation}{C.\arabic{equation}}

\begin{proof}
	 Let $(\alpha_v, \theta_v, I_v)$  denote a general strategy and $((\alpha_v)^*, (\theta_v)^*, (I_v)^*)$ denote the optimal strategy in the artificial market $\mathcal{M}_v$. Then, equation~\eqref{eq:wealth_fixed_v} gives the associated wealth $W_{v,t}$ under market $\mathcal{M}_v$
	\begin{equation*}
	 W^R_{v,t}=E^{Q_v}\bigg[\int_t^T {_{s-t}p_{x+t}}
	 \frac{\beta_{v,s}}{\beta_{v,t}}\,
	 [\lambda_{x+s}M^R_{v,s}-C_s^R-\delta(\widetilde v(s))],ds +{_{T-t}p_{x+t}}\frac{\beta_{v,T}}{\beta_{v,t}}W^R_{v,T}\,\bigg|\mathcal F_t\bigg].
	\end{equation*}
	It is obvious that
	\begin{equation}
		H_{v,t} = {_{t}p_{x}} \beta_{v,t}  W^R_{v,t} + \int_0^t {_{s}p_{x}}\beta_{v,s}  [\lambda_{x+s}M^R_{v,s}-C^R_s -\delta(\widetilde{v}(s))] ds \label{appendix_8_eq1}
	\end{equation}
	is a $Q_v$-martingale for $v \in \mathcal{N}^*$. By martingale representation theorem, there exists a $\mathbb{R}^4$-valued process $\widetilde{\Psi}_v$ with $\int_0^T ||\widetilde{\Psi}_{v,t}||^2 dt<\infty$, such that
	\begin{equation}
		H_{v,t} = W_{v,0} + \int_0^t \widetilde{\Psi}^{\top}_{v,s} dZ_{v,s}.\label{appendix_8_eq2}
	\end{equation}
	 Plug \eqref{appendix_8_eq2} into \eqref{appendix_8_eq1}, we drive
	\begin{align*}
		&W^R_{v,t} = ({_{t}p_{x}})^{-1}\beta^{-1}_{v,t} \left\{ H_{v,t} - \int_0^t {_{s}p_{x}} \beta_{v,s} [\lambda_{x+s}M^R_{v,s}-C^R_s -\delta(\widetilde{v}(s))] ds \right \}\\
		&=({_{t}p_{x}})^{-1}\beta^{-1}_{v,t} \left\{  W^R_{v,0} + \int_0^t \widetilde{\Psi}^{\top}_{v,s} dZ_{v,s} - \int_0^t {_{s}p_{x}} \beta_{v,s} [\lambda_{x+s}M^R_{v,s}-C^R_s -\delta(\widetilde{v}(s))] ds \right \}.
	\end{align*}
	Applying Itô's formula and the change of measure \eqref{eq:Zv}, we obtain
	\begin{align}
		&dW^R_{v,t} = (r_t+v_{0}(t) +\lambda_{x+t}) W^R_{v,t} dt + ({_{t}p_{x}})^{-1}\beta^{-1}_{v,t}\widetilde{\Psi}^{\top}_{v,t}[dZ_t + (\Lambda_t+v_{-}(t)-\sigma_{\Pi}) dt] \notag\\
		& - [\lambda_{x+t}M^R_{v,t} -C^R_t -\delta(\widetilde{v}(t))] dt.\label{appendix_8_eq3}
	\end{align}
	Choosing $({_{t}p_{x}})^{-1}\beta^{-1}_{v,t}\widetilde{\psi}^{\top}_{v,t}= (\theta^R_{v,t})^{\top}\Sigma-W^R_{v,t}\sigma^{\top}_{\Pi}$ and rewrite $M^R_{v,t} = W^R_{v,t} + I^R_{v,t}/\lambda_{x+t}$, then \eqref{appendix_8_eq3} can be rearranged as
	\begin{eqnarray}\label{appendix_8_wealth_SDE_2}
		dW^R_{v,t} &=& \left\{r_tW^R_{v,t} +[(\theta^R_{v,t})^{\top}\Sigma-W^R_{v,t}\sigma^{\top}_{\Pi}](\Lambda_t -\sigma_{\Pi}) \right\}dt + [(\theta^R_{v,t})^{\top}\Sigma-W^R_{v,t}\sigma^{\top}_{\Pi}]dZ_t\\
        &&+ [\alpha^R_{v,t} \widetilde{v}_{0}(t) + (\theta^R_{v,t})^{\top} \widetilde{v}_{-}(t)+\delta(\widetilde{v}(t))] dt
		  +C^R_t dt - I^R_{v,t}dt ,\notag\\
		W^R_{v,0} &=& w^R_0, (\alpha_v,\theta_v) \in \mathbb{R}^5.\notag
	\end{eqnarray} 
	which has no free disposal. The transform from $v(t)=(v_0(t),v_-(t))$ to $\widetilde{v}(t)=(\widetilde{v}_0(t),\widetilde{v}_-(t))$ follows the same argument in Appendix \ref{appendix_1}. Next, we extend the domain of $(\alpha^R_v,\theta^R_v)$ from the convex set $A$ to the whole domain $\mathbb{R}^5$ in order to find an explicit upper bound. By the definition \eqref{eq:support_function}, we have $\alpha^R_{v,t}\widetilde{v}_{0}(t) + (\theta^R_{v,t})^{\top}\widetilde{v}_{-}(t)  + \delta(\widetilde{v}(t))\geq 0$ for $(\alpha^R_v,\theta^R_v) \in A$. Consequently, the wealth process \eqref{appendix_8_wealth_SDE_2} has a drift term that is no smaller than that of the original wealth process \eqref{eq:real_wealth} almost surely for $(\alpha^R_v,\theta^R_v) \in A$. Since $A \subset \mathbb{R}^5$, maximizing the objective function $J(M^R_v,W^R_{v,T})$ under the wealth process \eqref{appendix_8_wealth_SDE_2} with a enlarged admissible set $(\alpha^R_v,\theta^R_v) \in \mathbb{R}^5$ provides an upper bound for the primal problem under the wealth process \eqref{eq:real_wealth} with the smaller admissible set $(\alpha^R_v,\theta^R_v) \in A$. Equivalently, the expected utility of an unconstrained investor who invests freely following \eqref{appendix_8_wealth_SDE_2} under artificial market $\mathcal{M}_v$ provides an upper bound for the primal problem \ref{primal_problem}. For $t\in [0,T]$, SDE \eqref{appendix_8_wealth_SDE_2} equals
	\begin{eqnarray*}
		dW^R_{v,t} 
		&=&\{[r_t+v_{0}(t)+\lambda_{x+t}]W^R_{v,t} + [(\theta^R_{v,t})^{\top}\Sigma-W^R_{v,t}\sigma^{\top}_{\Pi}](\Lambda_t + v_-(t) -\sigma_{\Pi})\}dt\\
        && + [(\theta^R_{v,t})^{\top}\Sigma-W^R_{v,t}\sigma^{\top}_{\Pi}] dZ_t  - [ \lambda_{x+t} M_{v,t} - C^R_t - \delta(\widetilde{v}(t))] dt.
	\end{eqnarray*}
	Define the value function $\widetilde{J}^S(t,W^R_{v,t},X_t;v)$ as
	\begin{align*}
		&\widetilde{J}^S(t,W^R_{v,t},X_t;v) = \sup \limits_{\theta^R_v,M^R_v} E_t\left[\int_t^T  {_{s-t}p_{x+t}}\lambda_{x+s} \frac{(M^R_{v,s})^{1-\gamma}}{1-\gamma} ds + {_{T-t}p_{x+t}} \frac{(W^R_{v,T})^{1-\gamma}}{1-\gamma}\right].
	\end{align*}
	Then, the corresponding HJB equation is given by
    \begin{align}
        &0=\sup \limits_{M^R_{v,t}}\left\{\lambda_{x+t}\frac{(M^R_{v,t})^{1-\gamma}}{1-\gamma}-\lambda_{x+t}M^R_{v,t}\frac{\partial \widetilde{J}^{S}}{\partial W_v}\right\}+\sup \limits_{\theta^R_{v,t}}\left\{\frac{\partial \widetilde{J}^{S}}{\partial W^R_v}(\theta^R_{v,t})^{\top} \Sigma [\Lambda_t + v_{-}(t)-\sigma_{\Pi}]\right.\notag\\
        &\left.+ \frac{1}{2}\frac{\partial^2 \widetilde{J}^{S}}{(\partial W^R_v)^2} [(\theta^R_{v,t})^{\top}\Sigma \Sigma^{\top} \theta^R_{v,t} - 2 W^R_{v,t}(\theta^R_{v,t})^{\top} \Sigma\sigma_{\Pi}]+ (\theta^R_{v,t})^{\top}\Sigma  (\Sigma_X)^{\top}\frac{\partial^2 \widetilde{J}^{S}}{\partial W^R_v \partial X^{\top}} \right\}\notag\\
        &+ \frac{\partial \widetilde{J}^{S}}{\partial t}-\lambda_{x+t}\widetilde{J}^{S}  + \frac{\partial \widetilde{J}^{S}}{\partial W^R_v}\{W^R_{v,t}[r_t+\lambda_{x+t}+v_{0}(t)-\sigma^{\top}_{\Pi}(\Lambda_t+v_{-}(t)-\sigma_{\Pi})] + C^R_t+\delta(\widetilde{v}(t))\}\notag\\
        &-\frac{\partial \widetilde{J}^{S}}{\partial X}K_X X_t + \frac{1}{2}\frac{\partial^2 \widetilde{J}^{S}}{(\partial W^R_v)^2}(W^R_{v,t})^2\sigma^{\top}_{\Pi}\sigma_{\Pi}+\frac{1}{2}\text{Tr}\left\{\Sigma^{\top}_X \frac{\partial^2 \widetilde{J}^{S}}{\partial X^{\top} \partial X}\Sigma_X\right\}-\frac{\partial^2 \widetilde{J}^{S}}{\partial W^R_v \partial X}\Sigma_X \sigma_{\Pi}W^R_{v,t}, \label{appendix_8_HJB_equation}
    \end{align}
		and the optimal unconstrained strategies are
		\begin{align}
        &(M^R_{v,t})^*=\left(\frac{\partial \widetilde{J}^{S}}{\partial W^R_v}\right)^{-\frac{1}{\gamma}},\label{appendix_8_M}\\
	    &(y_{v,t}^S)^*=
	    \frac{(\Sigma^{\top})^{-1}}
	    {\frac{\partial^2 \widetilde{J}^{S}}{(\partial W^R_v)^2}}
	    \left[
	    \frac{\partial^2 \widetilde{J}^{S}}{(\partial W^R_v)^2}
	    W^R_{v,t}\sigma_{\Pi}
	    +\frac{\partial \widetilde{J}^{S}}{\partial W^R_v}\kappa_{v,t}
	    -\Sigma^{\top}_X
	    \frac{\partial^2\widetilde{J}^{S}}
	    {\partial W^R_v\partial X^{\top}}
	    \right]. \label{appendix_8_theta}
	   \end{align}
    Moreover, the insurance strategy is
    $(I^R_{v,t})^*=\lambda_{x+t}[(M^R_{v,t})^*-W^R_{v,t}]$.

	Plugging the explicit solution \eqref{upper_bound_JR} into \eqref{appendix_8_HJB_equation}, one can verify that the value function $\widetilde{J}^S(t,W_{v,t};v)$ solves the HJB equation. Substituting
	\eqref{upper_bound_JR} and $v^*$ into \eqref{appendix_8_M} and
	\eqref{appendix_8_theta}  yields the corresponding candidate strategies \eqref{eq:qp_unconstrained_candidate} and \eqref{eq:qp_unconstrained_candidate_insurance}. The portfolio recovery is determined by \eqref{eq:method1_hamiltonian_qp}.

	 We now derive the Hamiltonian for policy recovery from the wealth dynamics in \eqref{eq:real_wealth}. Fix the dual process and evaluate the smooth value-function approximation at $(t,W_t^R,X_t)$. The unconstrained original-market Hamiltonian (the upper bound under $v \equiv 0$) is
	\begin{equation}
	\mathcal H_t^{S,\theta}(\theta)
	= \frac{\partial \widetilde{J}^S}{\partial W^R_v}\theta^\top\Sigma
	(\Lambda_t-\sigma_\Pi)
	+\theta^\top\Sigma\Sigma_X^\top
	\frac{\partial^2 \widetilde {J}^S}{\partial W^R_v \partial X^{\top}} +\frac{1}{2} \frac{\partial^2 \widetilde{J}^S}{\partial^2 W^R_v}
	\left[
	\theta^\top\Sigma\Sigma^\top\theta
	-2W_{v,t}^R\theta^\top\Sigma\sigma_\Pi
	\right].
	\label{eq:method1_proof_hamiltonian}
	\end{equation}
Here, we use the unconstrained Hamiltonian in the original market because the lower bound is computed in the original market $v \equiv 0$. One can also use the artificial-market Hamiltonian under $v^*$, but it yields a larger duality gap in the numerical example. Assume $\Sigma$ is nonsingular, $\gamma>0$ with $\gamma\ne1$, $\widetilde F_1^S>0$, $\widetilde F_2^S>0$. For the CRRA approximation,
\[
\frac{\partial \widetilde{J}^S}{\partial W^R_v}=(\widetilde F_1^S)^{-\gamma}(\widetilde F_2^S)^\gamma>0,
\qquad
 \frac{\partial^2 \widetilde{J}^S}{\partial^2 W^R_v}=-\gamma(\widetilde F_1^S)^{-\gamma-1}(\widetilde F_2^S)^\gamma<0.
\]
	Let $A_\Sigma=\Sigma\Sigma^\top$ and collect its terms linear in
	$\theta$ as
	\[
	b_t^S
	=\frac{\partial \widetilde{J}^S}{\partial W^R_v}\Sigma(\Lambda_t-\sigma_\Pi)
	+\Sigma\Sigma_X^\top\frac{\partial^2 \widetilde {J}^S}{\partial W^R_v \partial X^{\top}}
	-\frac{\partial^2 \widetilde{J}^S}{\partial^2 W^R_v} W_t^R\Sigma\sigma_\Pi.
	\]
	Then
	\[
	\mathcal H_t^{S,\theta}(\theta)
	=\frac{1}{2}\frac{\partial^2 \widetilde{J}^S}{\partial^2 W^R_v}\theta^\top A_\Sigma\theta
	+\theta^\top b_t^S.
	\]
	 The negative definite Hessian $(\partial^2 \widetilde{J}^S/\partial^2 W^R_v) A_\Sigma$ gives a unique unconstrained maximizer. Its first-order condition is
	\[
	\nabla_\theta\mathcal H_t^{S,\theta}(z_t^S)
	=\frac{\partial^2 \widetilde{J}^S}{\partial^2 W^R_v} A_\Sigma z_t^S+b_t^S=0,
	\]
Solving this linear system gives $z_t^S=-(\partial^2 \widetilde{J}^S/\partial^2 W^R_v)^{-1}A_\Sigma^{-1}b_t^S$.
	So that
	$b_t^S=-(\partial^2 \widetilde{J}^S/\partial^2 W^R_v) A_\Sigma z_t^S$.
	Substitution gives
	\[
	\mathcal H_t^{S,\theta}(\theta)
	=\frac12\frac{\partial^2 \widetilde{J}^S}{\partial^2 W^R_v}\theta^\top A_\Sigma\theta
	-\frac{\partial^2 \widetilde{J}^S}{\partial^2 W^R_v}\theta^\top A_\Sigma z_t^S,
	\]
	and, at the unconstrained maximizer,
	\[
	\mathcal H_t^{S,\theta}(z_t^S)
	=-\frac12\frac{\partial^2 \widetilde{J}^S}{\partial^2 W^R_v}
	(z_t^S)^\top A_\Sigma z_t^S.
	\]
	Hence, completing the square, the Hamiltonian difference between the unconstrained maximizer $z^S_t$ and the arbitrary strategy $\theta$ is given by
	\begin{align}
	\mathcal H_t^{S,\theta}(z_t^S)
	-\mathcal H_t^{S,\theta}(\theta) 
	=-\frac12\frac{\partial^2 \widetilde{J}^S}{\partial^2 W^R_v}
	(\theta-z_t^S)^\top
	A_\Sigma(\theta-z_t^S).
	\label{eq:method1_proof_hamiltonian_loss}
	\end{align}
	Since $\partial^2 \widetilde{J}^S/\partial^2 W^R_v<0$ and $A_\Sigma$ is positive definite,
	the expression in \eqref{eq:method1_proof_hamiltonian_loss} is
	nonnegative. Next, we replace $z^S_t$ with the candidate strategy $(y^S_{v^*,t})^*$ induced from the upper bound, as $(y^S_{v^*,t})^*$ is a good approximation of $z^S_t$. Then, we recover a $\theta$ close to our candidate strategy $(y^S_{v^*,t})^*$ in the Hamiltonian sense:
	\begin{align*}
	\arg\min \limits_{\substack{\theta\in \mathcal K(W_t^R)\\ \theta\geq0}}
	\left\{
	\mathcal H_t^{S,\theta}((y^S_{v^*,t})^*)
	-\mathcal H_t^{S,\theta}(\theta)
	\right\}=
	\arg\min \limits_{\substack{\theta\in \mathcal K(W_t^R)\\ \theta\geq0}}
	[\theta-(y^S_{v^*,t})^*]^\top\Sigma\Sigma^\top
	[\theta-(y^S_{v^*,t})^*],
	\end{align*}
	 where $\mathcal K$ is A-feasible set restricted to $\theta$. The same calculation also applies to the proposed nested Monte Carlo recovery using the portfolio candidate in Algorithm~\ref{alg:nmc_lower_bound}.

Here, the A-feasible set $\mathcal K$ is defined as
\[
\mathcal K(w)=\{\theta\in\mathbb R^4:\theta\geq0,\ \iota_4^\top\theta\leq w\},\qquad w\geq0.
\]
It is nonempty and compact. Since $A_\Sigma$ is positive definite, the projection objective is strictly convex, so the QP has a unique solution $\widehat\theta$. The additional upper limits $\theta_j\leq w$ in \eqref{eq:method1_hamiltonian_qp} are redundant. For $w>0$, Slater's condition holds, and the KKT conditions for the objective $\frac12(\theta-y)^\top A_\Sigma(\theta-y)$ are necessary and sufficient: There exist $\mu\in\mathbb R_+^4$ and $\eta\geq0$ such that
\begin{align*}
&A_\Sigma(\widehat\theta-y)-\mu+\eta\iota_4=0,\\
&\widehat\theta\geq0,\quad \iota_4^\top\widehat\theta\leq w,\quad
\mu_j\widehat\theta_j=0,\quad \eta(\iota_4^\top\widehat\theta-w)=0.
\end{align*}
At $w=0$ the feasible set is the singleton $\{0\}$, so the unique solution is $\widehat\theta=0$. Equivalently, for every $\theta\in\mathcal K(w)$ the solution satisfies
\[
(\theta-\widehat\theta)^\top A_\Sigma(\widehat\theta-y)\geq0.
\]

\end{proof}	

\section{Proof of Proposition \ref{semi_explicit_upperbound}}
\begin{proof}
According to \eqref{lag_upperbound}, maximizing the Lagrangian $\mathcal{L}$ with $(M_v^R, W^R_{v,T})$ and minimizing $(\Psi, v)$ provides an upper bound for the primal problem.  We define the following Lagrangian, a linear combination of the objective and the static budget constraint, to obtain an upper bound at each time $t$.
\begin{align*}
    &\mathcal{L}(t, M_v^R, W^R_{v,T}, \psi, v) = E_t\left[\int_t^T  {_{s-t}p_{x+t}}\lambda_{x+s} \frac{(M^R_{v,s})^{1-\gamma}}{1-\gamma} ds + {_{T-t}p_{x+t}}\frac{(W^R_{v,T})^{1-\gamma}}{1-\gamma} \right] \\
    &+ \psi \left\{W^R_{v,t} + E_t\left[\int_t^T {_{s-t}p_{x+t}}\frac{\pi_{v,s}}{\pi_{v,t}}[C^R_{s}+\delta(\widetilde{v}_s)-\lambda_{x+s}M^R_{v,s}]ds - {_{T-t}p_{x+t}}\frac{\pi_{v,T}}{\pi_{v,t}}W^R_{v,T}\right]\right\},
\end{align*}
where $\psi>0$ is the Lagrangian multiplier, and $\delta(\widetilde{v})=0$ for the  portfolio-mix constraint. By the first-order condition, we can obtain the optimal candidate strategies
\begin{eqnarray*}
	\left\{
	\begin{array}{l}
		(M^R_{v,s})^*=\psi^{-\frac{1}{\gamma}}\left(\frac{\pi_{v,s}}{\pi_{v,t}}\right)^{-\frac{1}{\gamma}},\\
		(W^R_{v,T})^*=\psi^{-\frac{1}{\gamma}}\left(\frac{\pi_{v,T}}{\pi_{v,t}}\right)^{-\frac{1}{\gamma}}.
	\end{array}
	\right.
\end{eqnarray*}
 Substituting these strategies into the static budget constraint yields the optimal Lagrangian multiplier
\begin{eqnarray*}
    \psi^* &=& \left\{ \int_t^T {_{s-t}p_{x+t}} \lambda_{x+s} E_t\left[\left(\frac{\pi_{v,s}}{\pi_{v,t}}\right)^{-\frac{1-\gamma}{\gamma}}\right]ds + {_{T-t}p_{x+t}}  E_t\left[\left(\frac{\pi_{v,T}}{\pi_{v,t}}\right)^{-\frac{1-\gamma}{\gamma}}\right]\right\}^{\gamma} \\
    &&\left\{W^R_{v,t}+ \int_t^T {_{s-t}p_{x+t}}E_t\left[\frac{\pi_{v,s}}{\pi_{v,t}}\right][C^R_s+\delta(\widetilde v_s)] ds \right\}^{-\gamma}.
\end{eqnarray*}
Since $\widetilde v_t=\mathcal{N\!N}_{\eta}(t,X_t)$ or $(u_t,s_t)=\mathcal{N\!N}_{\eta}(t,X_t)$, together with
\begin{eqnarray*}
    \left(\frac{\pi_{v,s}}{\pi_{v,t}}\right)^{-\frac{1-\gamma}{\gamma}}=\exp\left\{\frac{1-\gamma}{\gamma}\int_t^s r_u + v_{0,u} du + \frac{1}{2}\frac{1-\gamma}{\gamma}\int_t^s \kappa^2_{v,u}du -\frac{1-\gamma}{\gamma}\int_t^s\kappa_{v,u}dZ_u\right\},
\end{eqnarray*}
we obtain that the conditional expectation $E_t\left[\left(\pi_{v,s}/\pi_{v,t}\right)^{-\frac{1-\gamma}{\gamma}}\right]$ is a function of $(t,X_t, s)$.
Finally,  substituting the optimal $((M^R_{v,s})^*, (W^R_{v,T})^*)$ and $\psi^*$ into the objective, we obtain the value function \eqref{J_NN}. By the dynamic programming principle, we can derive the HJB equation under the artificial market
\begin{eqnarray*}
    0&=&\sup \limits_{M^R_{v,t}}\left\{\lambda_{x+t}\frac{(M^R_{v,t})^{1-\gamma}}{1-\gamma}-\lambda_{x+t}M^R_{v,t}\frac{\partial \widetilde{J}^{NN}}{\partial W_v}\right\}+\sup \limits_{\theta^R_{v,t}}\left\{\frac{\partial \widetilde{J}^{NN}}{\partial W^R_v}(\theta^R_{v,t})^{\top} \Sigma [\Lambda_t + v_{-,t}-\sigma_{\Pi}]\right.\\
    &&\left.+ \frac{1}{2}\frac{\partial^2 \widetilde{J}^{NN}}{(\partial W^R_v)^2} [(\theta^R_{v,t})^{\top}\Sigma \Sigma^{\top} \theta^R_{v,t} - 2 W^R_{v,t}(\theta^R_{v,t})^{\top} \Sigma\sigma_{\Pi}]+ (\theta^R_{v,t})^{\top}\Sigma  (\Sigma_X)^{\top}\frac{\partial^2 \widetilde{J}^{NN}}{\partial W^R_v \partial X^{\top}} \right\}\\
    &&+ \frac{\partial \widetilde{J}^{NN}}{\partial t}-\lambda_{x+t}\widetilde{J}^{NN}  + \frac{\partial \widetilde{J}^{NN}}{\partial W^R_v}\{W^R_{v,t}[r_t+\lambda_{x+t}+v_{0,t}-\sigma^{\top}_{\Pi}(\Lambda_t+v_{-,t}-\sigma_{\Pi})] + C^R_t+\delta(\widetilde{v}_t)\}\\
    &&-\frac{\partial \widetilde{J}^{NN}}{\partial X}K_X X_t + \frac{1}{2}\frac{\partial^2 \widetilde{J}^{NN}}{(\partial W^R_v)^2}(W^R_{v,t})^2\sigma^{\top}_{\Pi}\sigma_{\Pi}+\frac{1}{2}\text{Tr}\left\{\Sigma^{\top}_X \frac{\partial^2 \widetilde{J}^{NN}}{\partial X^{\top} \partial X}\Sigma_X\right\}-\frac{\partial^2 \widetilde{J}^{NN}}{\partial W^R_v \partial X}\Sigma_X \sigma_{\Pi}W^R_{v,t},
\end{eqnarray*}
where $\text{Tr}$ is the trace of a matrix. Solving the HJB equation for the optimal unconstrained strategies, we have
\begin{eqnarray*}
	\left\{
	\begin{array}{l}
    (M^R_{v,t})^*=\left(\frac{\partial \widetilde{J}^{NN}}{\partial W^R_v}\right)^{-\frac{1}{\gamma}},\\
		(y^{NN}_{v,t})^*= \frac{(\Sigma^{\top})^{-1}}{\frac{\partial^2 \widetilde{J}^{NN}}{(\partial W^R_v)^2}}\left[\frac{\partial^2 \widetilde{J}^{NN}}{(\partial W^R_v)^2}W^R_{v,t}\sigma_{\Pi} + \frac{\partial \widetilde{J}^{NN}}{\partial W^R_v}\kappa_{v,t} - \Sigma^{\top}_X \frac{\partial^2\widetilde{J}^{NN}}{\partial W^R_v\partial X^{\top}}   \right],
	\end{array}
	\right.
\end{eqnarray*}
where the insurance strategy is
$(I^R_{v,t})^*=\lambda_{x+t}[(M^R_{v,t})^*-W^R_{v,t}]$.
Substituting \eqref{J_NN} and $v^*$ into these candidates gives
\eqref{J_NN_strategy_theta} and \eqref{J_NN_strategy_I}.  For the portfolio recovery, we can obtain the candidate strategy by substituting  \eqref{J_NN_strategy_theta} into the Hamiltonian first-order condition and applying the QP algorithm, same as the proof of Proposition~\ref{explicit_solution_1}.
\end{proof}

\end{document}